\documentclass[journal]{IEEEtran}
\makeatletter
\long\def\@makecaption#1#2{\ifx\@captype\@IEEEtablestring%
\footnotesize\begin{center}{\normalfont\footnotesize #1}\\
{\normalfont\footnotesize\scshape #2}\end{center}%
\@IEEEtablecaptionsepspace
\else
\@IEEEfigurecaptionsepspace
\setbox\@tempboxa\hbox{\normalfont\footnotesize {#1.}~~ #2}%
\ifdim \wd\@tempboxa >\hsize%
\setbox\@tempboxa\hbox{\normalfont\footnotesize {#1.}~~ }%
\parbox[t]{\hsize}{\normalfont\footnotesize \noindent\unhbox\@tempboxa#2}%
\else
\hbox to\hsize{\normalfont\footnotesize\hfil\box\@tempboxa\hfil}\fi\fi}

\usepackage[compatibility=false]{caption}
\DeclareCaptionFont{quackfont}{\fontfamily{ptm}\fontsize{7.5pt}{9pt}\selectfont}
\usepackage[labelformat=simple,font=quackfont]{subcaption}

\makeatother
\usepackage{lipsum}
\makeatletter
\let\MYcaption\@makecaption
\usepackage{cite}
\usepackage{graphicx}
\usepackage{subcaption}
\usepackage{float}
\usepackage{refstyle}
\usepackage{multicol, blindtext}
\usepackage{amsmath,amssymb,amsfonts}
\usepackage{adjustbox}
\newcommand{\norm}[1]{\left\lVert#1\right\rVert}

\usepackage[font=scriptsize]{caption}
\usepackage[font=footnotesize]{caption}
\usepackage{stfloats}
\usepackage{lmodern,bm}                
\usepackage[T1]{sansmath}
\usepackage{xcolor}
\usepackage{algorithm, algorithmic}
\usepackage{enumitem}
\usepackage{dutchcal}

\begin{document}

\title{{ A Memory-Efficient Learning Framework for Symbol Level Precoding with Quantized NN Weights}}

\author{Abdullahi~Mohammad,~\IEEEmembership{Student Member,~IEEE,}
        Christos~Masouros,~\IEEEmembership{Senior Member,~IEEE,}
        and~Yiannis~Andreopoulos,~\IEEEmembership{Senior Member,~IEEE}}

\maketitle

\begin{abstract}
This paper proposes a memory-efficient deep neural network (DNN) framework-based symbol level precoding (SLP). We focus on a DNN with realistic finite precision weights and adopt an unsupervised deep learning (DL) based SLP model (SLP-DNet). {We} apply a stochastic quantization (SQ) technique to obtain its corresponding quantized version called SLP-SQDNet. The proposed scheme {offers a scalable performance vs memory tradeoff, by quantizing a scalable percentage of the DNN weights, and we explore binary and ternary quantizations}. Our results show that while SLP-DNet provides near-optimal performance, its quantized versions through SQ yield $\sim3.46\times$ and $\sim2.64\times$ model compression for binary-based and ternary-based SLP-SQDNets, respectively. We also find that our proposals {offer $\sim20\times$ and $\sim10\times$ computational complexity reductions compared to SLP optimization-based and SLP-DNet, respectively.}
\end{abstract}


\IEEEpeerreviewmaketitle

\section{Introduction}
\IEEEPARstart{P}{recoding} using the known channel state information (CSI) at the transmitter has been proven to be an efficient interference management technique in a downlink multiuser multiple-input-single-output (MU-MISO) communication system {\cite{windpassinger2004precoding}\cite{stankovic2004multi}.} The precoding also enables many complex signal processing at the base station (BS), which simplifies users’ terminals. Classical block-level precoding (BLP) schemes, where the precoding coefficients are applied across a block of symbols (codewords), have proven to be less computationally expensive than the optimal dirty paper coding (DPC) but suffer performance degradation \cite{lee2007high}. {Masouros and Alsusa \cite{masouros2007novel} first proposed a method for classifying instantaneous interference into constructive and destructive. The suboptimal precoding strategies that exploit constructive interference (CI) were first introduced \cite{masouros2010correlation}.} Precoding methods based on optimization are appealing because of their amenability to achieve various performance targets. An optimization-based CI precoding was first introduced using a quadratic optimization strategy in light of vector perturbation precoding\cite{masouros2014vector}.\par {To further improve the performance, a precoding design termed symbol-level-precoding (SLP) that exploits the multiuser interference via CI with the known CSI and transforms it into useful power at the mobile user end has received a lot of attention\cite{alodeh2015constructive,masouros2015exploiting,amadori2016constant,spano2017symbol,masouros2018harvesting}.} The CI-based solution is suitable for practical implementation and has proven that {massive multiple-input-multiple-output (m-MIMO)} systems can take advantage of the CI with SLP \cite{li2018massive,khandaker2018constructive,li2020interference}.
The idea of CI combined with optimization has been applied in many wireless physical layer designs due to its performance gains over BLP schemes to achieve different objectives, such as transmit power minimization and SINR balancing problems \cite{li2015two,li2017exploiting,law2017transmit,li2017hybrid,li2018interference}. A closed-form precoding design with optimal performance for a CI exploitation in the MISO downlink for optimization with both strict and relaxed phase rotations was proposed by Li and Masouros \cite{li2018interference}. While CI-based precoding methods offer {superior} performance, computing them online on a symbol-by-symbol basis can be computationally demanding.\par
As a result of the proliferation of machine learning algorithms, the model-driven deep learning (DL) technique that exploits the expert's knowledge has been applied in many wireless communication problems due to its explicability, reliability, and low computational complexity \cite{samuel2019learning,mohammad2020complexity,liang2019deep}. Therefore, DL-based precoding designs that use domain knowledge have been recently proposed for MU-MISO downlink transmission \cite{alkhateeb2018deep,xia2019deep,de2018robust}. However, the drawback of such methods is that the optimization constraints are not directly integrated with the loss function. Furthermore, their performance is bounded by the assumptions and accuracy of the optimal solutions obtained from the optimization algorithm. An unsupervised deep unfolding precoding design termed ``SLP-DNet"\cite{mohammad2021unsupervised} that utilizes the specifics of the optimization objectives of the precoding problem has been proposed to address the issues mentioned above, {and will be used as our benchmark in this work.}\par
Typically, a DL model contains thousands or even millions of learnable parameters, usually stored in a 32-bit floating-point (FP32) numerical presentation, making the model computationally and memory {demanding} during inference and deployment. To facilitate the online training and deployment of a trained DL model at the device edge, light-weight deep neural network (DNN) designs with lower-precision numerical formats have gained significant attention within the deep learning community, {typically applied to image processing applications}\cite{hubara2017quantized,rastegari2016xnor,he2017channel,dong2019stochastic}. However, this concept has not been fully explored in wireless communications. In this work, we propose a DL model's structural simplification method through weights quantization for SLP design. We adopt the DL-based SLP model (SLP-DNet) introduced by Mohammad \textit{et al.} \cite{mohammad2021unsupervised}. Our contributions are summarized below:
\begin{itemize}
    \item {We propose a memory and complexity efficient DNN approach, applied} to the learning-based precoding framework (SLP-DNet) \cite{mohammad2021unsupervised}. {Specifically,} we propose an efficient model simplification via weights compression to accelerate both training and inference and facilitate deployment on the device edge.
    \item {We devise a scalable tradeoff between performance and inference complexity, by allowing a percentage of the DNN weights to be quantized, while retaining important weights in full-precision. By tuning the percentage of quantised weights, a scalable tradeoff between performance and complexity / memory efficiency is achieved.} 
    \item We {further} introduce a stochastic quantization (SQ) technique that uses the quantization error to {alleviate the loss} in performance caused by the nonhomogenous quantization errors of the conventional extreme quantization (binary and ternary). In the SQ technique, a fraction of the neural network (NN) weight matrix is quantized to lower {resolution} while the remaining is retained in its full-precision, resulting in a hybrid quantized weight matrix. The technique yields a memory-efficient DL-based SLP model with a good balance between the performance and the computational complexity. 
\end{itemize}
The remainder of the paper is structured as follows: System model and the review of the relevant precoding techniques are presented in Section \ref{section2}. We introduce a technique of designing compressed unsupervised learning-based SLP schemes in Section \ref{sq_dnn_design}. Simulations and results are presented in Section \ref{results}. Finally, Section \ref{conclusion} concludes the paper.\par
\textbf{Notations:} We use bold uppercase symbols for matrices, bold lowercase symbols for vectors and lowercase symbols for scalars. Operators $\norm{\cdot}_{2}$, $\norm{\cdot}_{1}$ and $|\cdot|$ denote $\mathbcal{l}_{2}$-\text{norm}, $\mathbcal{l}_{1}$-\text{norm} and absolute values, respectively. The symbol $\boldsymbol{\Omega}_{i}$ represents the \textit{i}-th trainable parameter associated with DNN layers. $\text{Re}\{\cdot\}$ and $\text{Im}\{\cdot\}$ represent real and imaginary parts of complex vector/matrix, respectively. Finally, notations $\mathbcal{L}(\cdot)$ and $\mathbcal{D}(\cdot)$ are used for the loss and parameter update functions, respectively.

\section{System Model and Symbol Level Precoding}\label{section2}
\subsection{System Model} 
Consider an MU-MISO downlink transmission in a single cell scenario where an $M$-antenna base BS serves $K$ single-antenna users. The data is transmitted to the users over flat-fading Rayleigh channel denoted by $\mathbf{h}_{i}\in \mathbb{C}^{M\times1}$. The received signal at the $i$-th user is expressed as \begin{equation}\label{received_signal}
y_{i}=\mathbf{h}^{T}_{i}\sum\limits ^{K}_{k=1}\mathbf{u}_{k}+n_{i},
\end{equation}
where $\mathbf{h}_{i}$, $\mathbf{u}_{i}$, ${n}_{i}$ represent the channel vector, precoding vector and additive white Gaussian noise for the \textit{i}-th user.

\subsection{Symbol Level Precoding Power Minimization}
\begin{figure}[!t]
    \includegraphics[width=2.8in,height=2.2in]{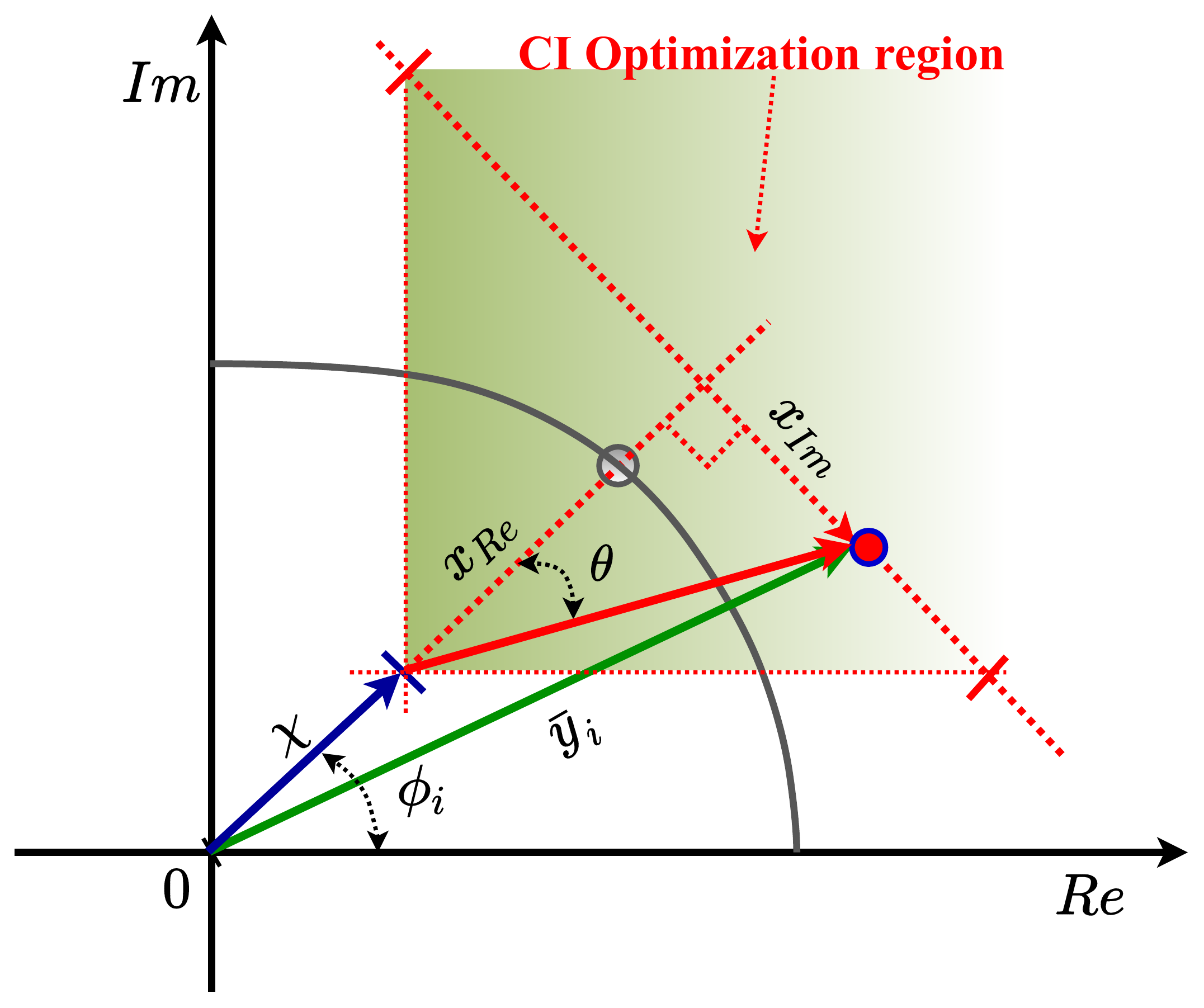}
    \caption{Generic geometrical optimization regions for interference exploitation \cite{masouros2015exploiting}}
    \label{fig:CI_GEOMETRY}
\end{figure}

The CI precoding scheme enhances the symbol detection by pushing the received signals away from the constellation detection boundaries without consuming extra transmission power \cite{masouros2015exploiting}. As an illustration, Fig. \ref{fig:CI_GEOMETRY} shows a symbolic example representing the constellation point $1+j$ in the QPSK. The green shaded area depicts the constructive region of the constellation based on the least distance $(\chi)$ from the decision boundaries, whose value is determined by the SNR constraints. This allows the interfering signals to align with the symbol of interest constructively through precoding vectors. We can observe that if the maximum angle shift in the CI region is zero, the interfering signals overlap entirely on the signal of interest ($\theta=0$), then the problem reduces to a strict phase angle optimization. It is important to note that the strict phase formulation is not appealing because it yields an increase in the transmission power compared to the corresponding relaxed version \cite{li2020tutorial}.
For simplicity, the following are defined according to \cite{masouros2015exploiting};
$\Tilde{\mathbf{h}}_{i}=\mathbf{h}_{i}\sum_{k=1}^{K}e^{j(\phi_{k}-\phi_{i})} \in{\mathbb{C}^{M\times1}}$, ${\mathbf{u}}=\sum_{k=1}^{K}\mathbf{u}_{k} \in{\mathbb{C}^{M\times1}}$, $\Tilde{\mathbf{h}}_{Ri}=\text{Re}(\Tilde{\mathbf{h}}_{i})$, $\Tilde{\mathbf{h}}_{Ii}=\text{Im}(\Tilde{\mathbf{h}}_{i})$, $\mathbf{u}_{R}=\text{Re}(\mathbf{u})$ and $\mathbf{u}_{I}=\text{Im}(\mathbf{u})$. Similarly, we also let $\boldsymbol{\Phi}_{i} =\begin{bmatrix}
\Tilde{\mathbf{h}}_{Ri}; \ \Tilde{\mathbf{h}}_{Ii}\
\end{bmatrix}$, $\mathbf{u}_{1} = [
\mathbf{u}_{R}\ \ -\mathbf{u}_{I}]^{T}$; where $
\boldsymbol{\Upsilon}=\begin{bmatrix}
\mathbf{O}_{M} & -\mathbf{I}_{M}\\
\mathbf{I}_{M} & \mathbf{O}_{M}
\end{bmatrix}\ \in \mathbb{R}^{2M \times 2M}.
$ {For the details and rational of the above definitions the reader is referred to [4].} We define the precoding and the channel matrices respectively as $\Tilde{\mathbf{H}}=[\Tilde{\mathbf{h}}_{1},\cdots,\Tilde{\mathbf{h}}_{K}]$ and $\mathbf{U}=[\mathbf{u}_{1},\cdots,\mathbf{u}_{K}]$. 
Therefore, for an $\mathbb{M}$-phase shift keying ($\mathbb{M}$-PSK) modulation scheme, where $\mathbb{M}$ is the modulation index, the optimization-based SLP for a nonrobust multicast power minimization is given by\cite{mohammad2021unsupervised}
\begin{equation}\label{P_relaxed2}
    \begin{aligned}
    & \underset{\mathbf{\{u_{1}\}}}{\text{min}}
    & & {\norm{\mathbf{u}_{1}}_{2}^2} \\
    & \text{s.t.}
    & & \bar{a} \leq {\boldsymbol{\Phi}_{i}^{T}\boldsymbol{\Upsilon}\mathbf{u}_{1}}\leq\bar{b}\ ,\ \forall {i}.
    \end{aligned}
\end{equation} 
where $\bar{a}=-\left(\boldsymbol{\Phi}_{i}^{T}\boldsymbol{\Upsilon}\mathbf{u}_{1}-\sqrt{\Gamma_{i}n_{0}}\right){\text{tan}{\theta}}$ and $\bar{b}=\left(\boldsymbol{\Phi}_{i}^{T}\boldsymbol{\Upsilon}\mathbf{u}_{1}-\sqrt{\Gamma_{i}n_{0}}\right){\text{tan}{\theta}}$, $\Gamma_{i}$ is the target SINR, $\theta=\pm\frac{\pi}{\mathbb{M}}$ is the maximum phase shift in the CI region.\par
{To avoid repetition, we} refer the reader to \cite{masouros2015exploiting}\cite{mohammad2021unsupervised} for details and the description of equivalent robust formulations under channel uncertainty. 
\subsection{Learning-Based SLP for Power Minimization {(SLP-DNet)}}\label{SLP}
This work is based on the unsupervised deep unfolding framework that unfolds the interior point method (IPM) \textit{`log'} barrier function based on the problem (\ref{P_relaxed2}) {by reformulating it as unconstrained subproblems per user expressed as
\begin{equation}\label{sub_prob}
    \begin{aligned}
    & \underset{\mathbf{u \in{\mathbb{R}^{2M\times1}}}}{\text{min}} f(\mathbf{u}_{1})+{\upsilon}{{B(\mathbf{u}_{1})}},
    \end{aligned}
\end{equation}}
{where $B(\mathbf{u}_{i})\triangleq\sum_{i=1}^{t}{\ln{\left(g(\mathbf{u}_{1i})\right)}}$ is the logarithmic barrier function and $\upsilon$ is the Lagrangian multiplier related to the inequality constraints. Here, the function, $g(\mathbf{u}_{i})$ is defined as $g(\mathbf{u}_{1})=\left(\boldsymbol{\Phi}_{i}^{T}\mathbf{u}_{1}-\sqrt{\Gamma_{i}n_{0}}\right)\text{tan}\theta-\left|{\boldsymbol{\Phi}_{i}^{T}\boldsymbol{\Upsilon}\mathbf{u}_{1}}\right|$ and $t$ is the number of the optimization variables.}\par
\begin{figure*}[!t]
\centering
    \includegraphics[width=7.2in,height=3.2in]{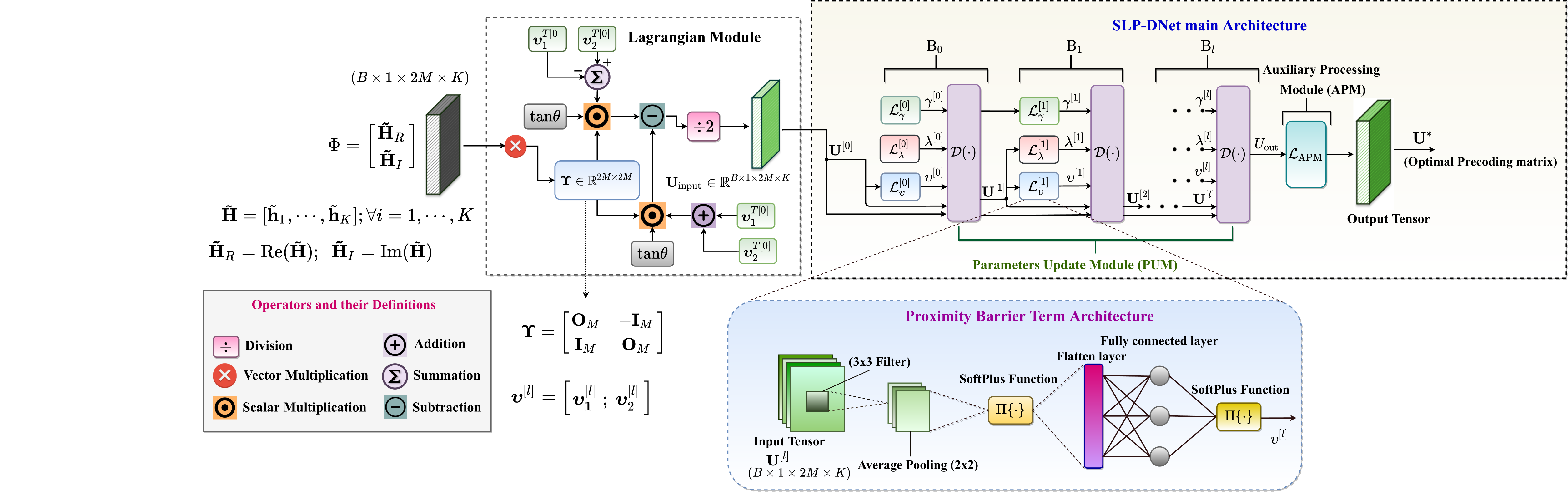}
    \caption{Learning-based symbol level precoding (SLP-DNet) Architecture \cite{mohammad2021unsupervised}}
    \label{fig:DNBF_Arch}
\end{figure*}

{To derive the learning architecture based on an IPM, we define a proximity barrier of (\ref{sub_prob}) as 
\begin{equation}\label{prox_op}
   \begin{aligned}
   \text{prox}_{\gamma{\upsilon}{B}}{(\mathbf{u}_{1})}= & \underset{\mathbf{u_{1} \in{\mathbb{R}^{2M\times{1}}}}}{\text{argmin}}
    & & {\frac{1}{2}\norm{\mathbf{u}_{0}-\mathbf{u}_{1}}}_{2}^{2}+\gamma{\upsilon}{B}({\mathbf{u}_{1}}), 
    \end{aligned}
\end{equation}
$\mathbf{u}_{0}$ is the initial precoding vector and $\gamma \in \{0,+\infty\}$ is the training step size.} The precoding vector for every \textit{l}-th iteration is obtained from the following learning update rule
\begin{equation}\label{beam_update_relaxed}
\mathbf{u}_{1}^{[l+1]}=\text{prox}_{\gamma^{[l]}\boldsymbol{\upsilon}^{[l]}B}\left(\mathbf{u}_{1}^{[l]}-\gamma^{[l]}\Delta{{H}(\mathbf{u}_{1}^{[l]},\lambda^{[l]})}\right)
\end{equation}
where ${H}(\mathbf{u}_{1}^{[l]},\lambda^{[l]})={\norm{\mathbf{u}_{1}}}_{2}^{2}+\lambda{\mathbf{u}_{1}}$, and $\Delta=\frac{\partial{{H}(\mathbf{u}_{1}^{[l]},\lambda^{l})}}{\partial{\mathbf{u}_{1}^{[l]}}}$. The parameter, $\lambda$ is introduced as an additional constraint to provide more stability to the learning architecture. Intuitively, NN cascade layers can be formed from (\ref{beam_update_relaxed}) as follows
\begin{equation}\label{beam_update2}
\mathbf{u}_{1}^{[l+1]}=\text{prox}_{\gamma^{[l]}{\upsilon}^{[l]}{B}}\left[\left(\mathbf{I}_{2M}-2\gamma^{[l]}\right)\mathbf{u}_{1}^{[r]}+\gamma^{[l]}\lambda^{[l]}\textbf{1}^{T}\right].
\end{equation}
where $\textbf{1}\in\mathbb{R}^{1\times2M}$ is a vector of ones. By letting $\mathbf{W}_{l}=\mathbf{I}_{2M}-2\gamma^{[l]}$, $\mathbf{b}_{l}=\gamma^{[l]}\lambda^{[l]}\textbf{1}^{T}$ and $\boldsymbol{\Pi}_{l}=\text{prox}_{\gamma^{[l]}{\upsilon}^{[l]}{B}}$, the \textit{l}-layer network $\mathcal{L}^{[l-1]}\cdots \mathcal{L}^{[0]}$ will correspond to the following
\begin{equation}\label{neural_net}
  \boldsymbol{\Pi}_{0}\left(\mathbf{W}_{0}+\mathbf{b}_{0}\right),\cdots,\boldsymbol{\Pi}_{l}\left(\mathbf{W}_{l}+\mathbf{b}_{l}\right)
\end{equation}
where $\mathbf{W}_{l}$ and $\mathbf{b}_{l}$ are described as weight and bias parameters respectively. The nonlinear activation functions are defined by $\boldsymbol{\Pi}_{l}$.
The SLP-DNet structure as shown in Fig. \ref{fig:DNBF_Arch} is built based on (\ref{beam_update2}) and the Algorithms 1 and 2 of \cite{mohammad2021unsupervised}. 
As shown in Fig. \ref{fig:DNBF_Arch}, SLP-DNet has two main units; the parameter update module (PUM) and the auxiliary processing module (APM). The PUM has three core components associated with Lagrangian multiplier ($\upsilon$), the auxiliary parameter ($\lambda$), and the training step-size ($\gamma$), which are updated based on the following
\begin{equation}\label{update_func}
\mathcal{D}(\mathbf{u}_{1},\boldsymbol{\upsilon},\gamma,\lambda)=\text{prox}_{\gamma^{[l]}{\upsilon}^{[l]}B}\left(\mathbf{u}_{1}^{[l]}-\gamma^{[l]}\Delta{{H}(\mathbf{u}_{1}^{[l]},\lambda^{[l]})}\right).
\end{equation}
The structure that is related to the inequality constraint {in (\ref{P_relaxed2})} is the proximity barrier term. It is constructed with one convolutional layer, an average pooling layer, a fully connected layer, and a softPlus layer to constrain the output to a positive real value to satisfy the inequality constraint.
\par 
The loss function over $N$ batch training samples (batch size or the number of channel realization) is Lagrangian function expressed as
\begin{multline}\label{Lag_relaxed2}
\mathcal{L}(\mathbf{u}_{1} ,\boldsymbol{\upsilon}_{1} ,\ \boldsymbol{\upsilon}_{2}) =\frac{1}{N}\sum^{N}_{i=1}\Vert \mathbf{u}_{1}\Vert_{2}^{2} \\
+\frac{1}{N}\sum^{N}_{i=1}\left(\boldsymbol{\upsilon}_{1}\left(\boldsymbol{\Phi}^{T}_{i}\boldsymbol{\Upsilon }\mathbf{u}_{1}-\boldsymbol{\Phi}_{i}^{T}\mathbf{u}_{1}\text{tan}{\theta}+\sqrt{\Gamma_{i}n_{0}}\right)\right)\\
-\frac{1}{N}\sum^{N}_{i=1}\left(\boldsymbol{\upsilon}_{2}\left(\boldsymbol{\Phi}^{T}_{i}\boldsymbol{\Upsilon }\mathbf{u}_{1}+\boldsymbol{\Phi}_{i}^{T}\mathbf{u}_{1}\text{tan}{\theta}-\sqrt{\Gamma_{i}n_{0}}\right)\right) \\
+\frac{\mu}{NL}\sum^{N}_{i=1}\sum_{l=1}^{L}\Vert \boldsymbol{\Omega}_{i}\Vert_{2}^{2},
\end{multline}
where $\boldsymbol{\Omega}_{i}$ are the trainable parameters of the \textit{l}-th layers associated with the weights and biases, and $\mu >0$ is the penalty parameter that controls the bias and variance of the trainable coefficients.\par
The optimal precoder is obtained from the Lagrangian function (\ref{Lag_relaxed2}) as 
\begin{equation}\label{optima_prec_rel}
\mathbf{u}_{1}=\frac{\left(\boldsymbol{\upsilon}_{1}^{T}+\boldsymbol{\upsilon}_{2}^{T}\right)\cdotp\boldsymbol{\Phi}_{i}\text{tan}{\theta}-\left(\boldsymbol{\upsilon}_{1}^{T}-\boldsymbol{\upsilon}_{2}^{T}\right)\cdotp\boldsymbol{\Upsilon^{T}}\boldsymbol{\Phi}_{i}\tan{\theta}}{2}.
\end{equation}

\subsection{Robust SLP-DNet}  
{In a similar fashion to the above, we can derive a CSI-robust SLP-DNet from the robust SLP formulation under worst-case CSI-error. The robust SLP is given by \cite{mohammad2021unsupervised}
\begin{equation} \label{robust_multi4}
    \scalebox{.94}{$\begin{aligned}
    & \underset{\mathbf{\{u_{2}\}}}{\text{min}}
    & & {\norm{\mathbf{u}_{2}}_{2}^2} \\
    & \text{s.t.}
    & & \boldsymbol{\Phi}^{T}\mathbf{Q}_{1}\mathbf{u}_{2}+\varsigma\norm{\mathbf{Q}_{1}\mathbf{u}_{2}}_{2}+\sqrt{\Gamma n_{0}}\text{tan}{\theta}\leq0\  \forall{i}\\ 
    &&& \boldsymbol{\Phi}^{T}\mathbf{Q}_{2}\mathbf{u}_{2}+\varsigma\norm{\mathbf{Q}_{2}\mathbf{u}_{2}}_{2}+\sqrt{\Gamma n_{0}}\text{tan}{\theta}\leq0\ 
    \forall{i}.
    \end{aligned}$}
\end{equation}
For convenience, we introduce new notations as follows: $\mathbf{Q}_{1}=\left(\boldsymbol{\Upsilon}-\mathbf{I}_{2M}\text{tan}{\theta}\right)$ and $\mathbf{Q}_{2}=\left(\boldsymbol{\Upsilon}+\mathbf{I}_{2M}\text{tan}{\theta}\right)$ and $\varsigma^{2}$ is the CSI error bound. (\ref{robust_multi4}) is a second order cone programming (SOCP) and can be solved using convex optimization software package.}\par
It is important to note that the structure of the robust SLP-DNet is obtained by following similar steps from (\ref{sub_prob})-(\ref{update_func}) of Subsection \ref{SLP} by transforming (\ref{robust_multi4}) to its equivalent unfolded IPM \textit{`log'} barrier form. The loss function is obtained from the Lagrangian of (\ref{robust_multi4}) as
\begin{multline}\label{Lag_robust_reg}
\mathcal{L}_{\text{robust}}(\mathbf{u}_{2},\ \boldsymbol{\upsilon}_{1},\ \boldsymbol{\upsilon}_{2}) =\frac{1}{N}\sum^{N}_{i=1}\Vert \mathbf{u}_{2}\Vert_{2}^{2} \\
+\frac{\boldsymbol{\upsilon}_{1}}{N}\sum^{N}_{i=1}\left({\varsigma}^{2}\norm{\mathbf{Q}_{1}\mathbf{u}_{2}}_{2}^{2}-\left(\sqrt{\Gamma n_{0}}\text{tan}{\theta}-\boldsymbol{\Phi}^{T}\mathbf{Q}_{1}\mathbf{u}_{2}\right)^{2}\right)\\
+\frac{\boldsymbol{\upsilon}_{2}}{N}\sum^{N}_{i=1}\left({\varsigma}^{2}\norm{\mathbf{Q}_{2}\mathbf{u}_{2}}_{2}^{2}-\left(\sqrt{\Gamma n_{0}}\text{tan}{\theta}-\boldsymbol{\Phi}^{T}\mathbf{Q}_{2}\mathbf{u}_{2}\right)^{2}\right)\\
+\frac{\mu}{NL}\sum^{N}_{i=1} \sum_{i=1}^{L}\Vert \boldsymbol{\Omega}_{i}\Vert_{2}^{2}.
\end{multline}
where $\begin{bmatrix}\| \mathbf{Q}_{1} \|_{2}^{2} & 
 \| \mathbf{Q}_{2} \|_{2}^{2}\end{bmatrix} =\bar{\mathbf{q}}_{\text{norm}}$, $\begin{bmatrix}\mathbf{Q}_{1} & \mathbf{Q}_{2} \end{bmatrix} =\Tilde{\mathbf{Q}}$ and $\begin{bmatrix}\boldsymbol{\upsilon}_{1} & \boldsymbol{\upsilon}_{2} \end{bmatrix} =\Tilde{\boldsymbol{\upsilon}}$.\par 
The optimal precoder can be easily obtained from (\ref{Lag_robust_reg})
\begin{equation}\label{robust_optimal}  
    \mathbf{u}_{2}=-\boldsymbol{\Phi}\mathbf{\Tilde{Q}}{\boldsymbol{\Tilde{\upsilon}}}^{T}{\mathbf{X}}^{-1}\sqrt{\Gamma{n}_{0}}{\text{tan}{\theta}},
\end{equation}
where ${\mathbf{X}}=\left(\mathbf{I}_{2M}+\mathbf{\Tilde{q}}_{\text{norm}}\boldsymbol{\Tilde{\upsilon}}^{T}\left({\varsigma}^{2}-\boldsymbol{\Phi^{T}\Phi}\right)\right)$. 
Note that the Lagrange multipliers $\boldsymbol{\upsilon}_{1}$ and $\boldsymbol{\upsilon}_{2}$ are associated with the barrier term and are randomly initialized from a uniform distribution.
{\section{{Preliminaries of NN Weight Quantization}}}\label{low-bit_NN}
Traditionally, DNN is designed with full-precision weights and activations. {This can result in significant memory consumption and computational complexity. For this reason, there has been a recent drive to reduce the DNN model size, driven from the image processing research \cite{hubara2017quantized}. DNN acceleration techniques can be broadly classified into three categories:}
\begin{enumerate}[label=\roman*.]
    \item Structured simplification: This involves a systematic approach of network factorization (factorizes a convolutional layer into many efficient ones), channel pruning, sparse connections to reduce the size of the {DNN} model \cite{mohammad2020accelerated}.
    \item Optimized Implementation: This approach uses Fast Fourier Transform (FFT) based on NVIDIA’s cuFFT library to provide significant speedups \cite{abtahi2017accelerating}.
    \item Quantization: In this technique, the computations involving weights, activations, and sometimes input tensors are performed at lower bit-widths than floating-point precision \cite{hubara2017quantized}.
\end{enumerate}
Among the above three model simplification techniques, quantization is most appealing because, in addition to model reduction, most MACs operations required to compute the neurons' weighted sums are replaced by simple binary operations (bit-wise or XNOR operations). Quantization improves both training and inference efficiencies; and reduces hardware requirements during model deployment on the edged-devices.\par
Typically, the weights of \textit{l}-th layer DNN architecture are represented by \cite{rastegari2016xnor} $\boldsymbol{\mathcal{W}}_{l} =\{\mathbf{W}_{i},\cdots ,\ \mathbf{W}_{m}\}$ for $\forall\ i=1,\ \cdots, m$, where $m$ is the number of kernels/filters (output channels). The \textit{n}-dimensional weight tensor $\mathbf{W}_{i} \in \mathbb{R}^{n}$, $n=c\times w\times h$ in \textit{l}-th convolutional layer, where $c\times w\times h$ represents the input channels, filter width and filter height respectively, and for a fully connected layer, $n=m\times c$ (number of the output and input neurons, respectively). For convenience, in what follows, we drop the {kernel subscript}.
\subsubsection{Binary Weights}
The real-valued weights are converted to $\left(\mathbf{B}_{w} \in \{+1, -1\}^{n}\right)$. A full-precision 32-bit weight matrix is binarized as follows\cite{courbariaux2015binaryconnect}
\begin{equation}\label{det_binary}
    \mathbf{B}_{w}=sign(\mathbf{W})=\begin{cases}
    +1 & \text{if} \ \mathbf{W} \geq 0\\
    -1 & \text{otherwise,}
    \end{cases}
\end{equation}
A more robust binarized {weight} ``BWN" is proposed as an extension of a straightforward binary network (Binary Connect) by introducing a real scaling factor $\beta \in \mathbb{R}^{+}$ such that $\mathbf{W}\approx\beta\mathbf{B}_{\text{w}}$ by solving an optimization problem \cite{rastegari2016xnor}    
\begin{equation}\label{binary_optim}
   \begin{aligned}
   J(\mathbf{B}_{\text{w}},\beta)=& \underset{(\mathbf{B}_{\text{w}},\beta)}{\text{argmin}}
    & & {\norm{\mathbf{W}-\beta\mathbf{B}_{\text{w}}}_{2}^{2}} \\
    \end{aligned},
\end{equation}
and this yields
\begin{equation}\label{binary_weight}
\begin{split}
    \mathbf{B}_{\text{w}}^{*}=sign(\mathbf{W})\\
    \beta^{*}=\frac{1}{n}\norm{\mathbf{W}}_{1}
\end{split}
\end{equation}

\subsubsection{Ternary Weights}
A ternary weighted network (TWN) is the one in which an extra 0 state is introduced into BWN to solve the following optimization problem \cite{alemdar2017ternary}
\begin{equation}\label{ternary_opt}
\begin{cases}
\beta ^{*} ,\mathbf{B}^{*}_{\mathbf{W}} = & \underset{\beta ,\ \mathbf{B}_{\text{w}}}{\text{argmin}} \ J(\mathbf{\beta ,\ B}_{\text{w}} \ ) =\| \mathbf{W} -\beta \mathbf{B}_{\text{w}} \|_{2} ^{2}\\
\text{s.t.} & \beta \geq 0,\ \mathbf{B}_{\text{w}} \in \{-1,\ 0,\ +1\}^{n},
\end{cases}
\end{equation}
and solving (\ref{ternary_opt}) gives
\begin{equation}\label{ternary_weight}
\mathbf{B}^{*}_{\text{w}} =\begin{cases}
+1 & \text{, if} \ \mathbf{W}  >\delta \\
0 & \text{, if} \ |\mathbf{W} |\leq \delta \\
-1 & \text{, }\text{if} \ \mathbf{W} < -\delta,
\end{cases}
\end{equation}
where $\delta =\frac{0.7}{n}\sum\limits ^{n}_{i=1} |\mathbf{W} |$ and $\beta ^{*} =\frac{1}{\mathbf{I}_{\delta }}\sum\limits _{i\in \mathbf{I}_{\delta}} |\mathbf{W} |$,\\ $\mathbf{I}_{\delta} = \{|\mathbf{W}|>\delta\}$ is the cardinality of set $\mathbf{I}_{\delta}$.

\section{Proposed Low-bit SLP-DNet Design}\label{sq_dnn_design}
\subsection{Low-Bit Weights and Stochastic Division}
The existing works on low-bit  DNNs design focus only on reducing the bit-widths of the weights and activations to speed up the training and inference times and also improve memory efficiency. However, in low-bit  DNNs designs, the impact of quantization on the performance of the learning algorithm has not been fully explored and understood. In this work, we adopt a quantization technique proposed in \cite{dong2019stochastic} and propose a simple linear probability function of selecting the filter weights to be quantized for designing a low-bit scalable learning-based precoder.\par 
The weight matrix of each layer of the DNN can be expressed as: $\mathbcal{W}=\{\mathbf{W}_{1},\cdots, \mathbf{W}_{n}\}$. Here, the rows of the weight matrix are partitioned into two parts according to the following
\begin{equation}\label{weight_partition1}
    \mathbcal{W}=\{\mathcal{W}_{q},\ \mathbcal{W}_{f}\},
\end{equation}  
where $\mathbcal{W}_{q}=\{\mathbf{W}_{q1},\ \cdots,\ \mathbf{W}_{qM}\}$ and $\mathbcal{W}_{f}=\{\mathbf{W}_{f1},\ \cdots,\ \mathbf{W}_{fN}\}$ represent the quantized and full-precision parts of the weight respectively, and should satisfy the condition below
\begin{equation}\label{weight_partition2}
    \mathbcal{W}=\mathcal{W}_{q} \cup \mathbcal{W}_{f}\ \text{and}\ \mathcal{W}_{q} \cap \mathbcal{W}_{f}=\emptyset.
\end{equation} 
As seen from (\ref{weight_partition1}), one {subset} of the weight $\mathbcal{W}_{q}$ is quantized to a low bit-width while the {remaining} $\mathbcal{W}_{f}$ is kept in its full-precision form, so that the entire weights matrix is composed of both binary and floating-point values. {Note that a fully quantized DNN can be obtained by setting $\mathbcal{W}_f$ to a null set}.\par
\begin{figure*}[!th]\label{lottery_block}
    \includegraphics[width=\linewidth]{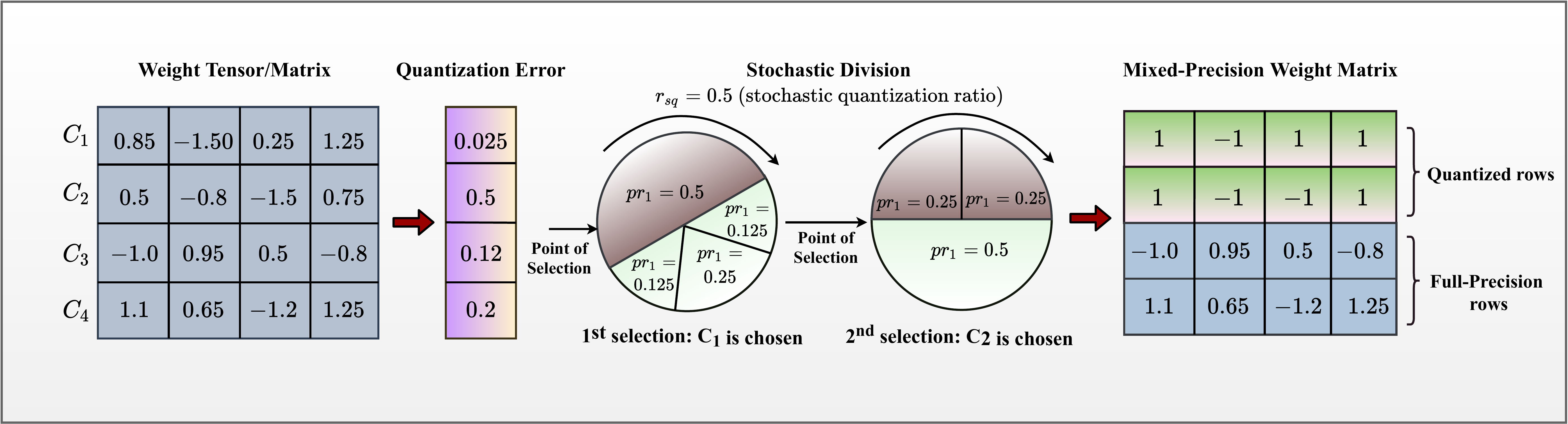}
    \caption{Stochastic Quantization Weight Matrix Partitioning Procedure}
    \label{fig:SQ_DISC_ALGO}
\end{figure*}

Suppose $r_{sq}$ is the quantization ratio (QR) {(i.e., the percentage of weights quantized as a fraction of the total weights in the DNN)}, and $n$ is the length of the weight matrix (number of elements), the number of elements in the quantization group is $M_{q}=r_{sq}n$ while that of a full-precision parts is $M_{f}=(1-r_{sq})n$. The QR {can be} gradually increased to 100\% until the entire network is finally quantized. To select the channel to be quantized, we adopt a lottery disc algorithm as in \cite{dong2019stochastic}. It can be observed in Fig. \ref{fig:SQ_DISC_ALGO} that each sector of the disc represents a probability of selecting a channel (row of weight matrix). The disc is rotated by choosing a value from the uniform distribution whose magnitude is slightly above the probability value. After every selection, the probability is reset (i.e., $p_{j}=0$) to ensure that a channel is selected without replacement as summarized in Algorithm \ref{algorithm_SQ_Division}.

\begin{algorithm}
 \caption{Circular Lottery Algorithm for Weight matrix Division}
 \begin{algorithmic}[1]\label{algorithm_SQ_Division}
 \renewcommand{\algorithmicrequire}{\textbf{Input:}}
 \renewcommand{\algorithmicensure}{\textbf{Output:}}
 \REQUIRE $r_{sq}$ Stochastic Quantization ratio and Weight matrix ($\mathbcal{W})$
 \ENSURE  $\mathbcal{W}_{q}$ and $\mathbcal{W}_{f}$
  \STATE \textit{Initialization}:\\
  $\mathbcal{W}_{q}=\mathbcal{W}_{r}=\emptyset$
  \STATE Compute QP function $\mathbf{pr}\in \mathbb{R}^{n} \forall{i}\ \{i=1,\ \cdots,\ n\}$ based on (\ref{q_error})
  \STATE $M_{q}=r_{sq}n$
   \FOR{$j=1$ to $M_{q}$}
        \STATE $\hat{\mathbf{p}}_{r}=\frac{\mathbf{p}_{r}}{\norm{\mathbf{p}_{r}}_{1}}$ (normalized probability)
    
  \STATE Select a random value $\vartheta_{j}\in\{0, 1\}$ from a random uniform distribution
  \STATE Set $s_{j}=0$ and $i=0$
  \WHILE{$s_{j}< \vartheta_{j}$} 
    \STATE $i=i+1$
    \STATE $s_{j}=s_{j}+\hat{p}_{r}$
  \ENDWHILE 
  \STATE Compute: $\mathbcal{W}_{q}=\mathbcal{W}_{q}\cup\mathbcal{W}$
  \STATE Reset $pr_{i}=0$ \COMMENT{This is to avoid \textit{i-th} channel weight from being selected again}
  \ENDFOR
  $\mathbcal{W}_{r}=\mathbcal{W}\setminus\mathbcal{W}_{q}$
 \end{algorithmic} 
\end{algorithm}

\subsection{Quantization Error and Quantization Probability} 
Recall that {classical} binarized DNNs {suffer} a significant performance loss due heterogeneous nature of the quantization error (QE) over the entire network. The performance can, however, be improved by stochastically selecting the filter or channel weight matrix to be quantized using a random probability distribution based on the QE between the real-valued and quantized weights as follows
\begin{equation}\label{q_error}
    e_{j}=\frac{\norm{\mathbf{W}_{j}-\mathbf{Q}_{j}^{*}}_{1}}{\norm{\mathbf{W}_{j}}};
\end{equation}
where $\mathbf{Q}_{j}^{*}$ could be binary or ternary based on (\ref{binary_weight}) or (\ref{ternary_weight}).\par

We define the vector of the \textit{n-th} row weight matrix of a given layer as $\mathbf{e}=[e_{1},\ \cdots,\ e_{n}]$. The quantization probability is formulated such that a higher probability is assigned to filter/weights if the quantization error is small because quantizing these weights does not yield a significant loss of accuracy or performance.   
For a given weight matrix, QR, and quantization probability (QP), a channel is randomly sampled without replacement using a circular lottery Algorithm \ref{algorithm_SQ_Division}. From this, we can observe that the QP function is inversely proportional to QE and is defined as $f_{p}=\frac{1}{e+\delta}$, where $\delta=10^{-6}$ to avoid possible numerical overflow. The QP function is monotonically non-decreasing to prioritize the selection of the channels/weights to be quantized. Different monotonically non-decreasing functions are:
\begin{itemize}
    \item Uniform function: $pr_{j}=\frac{1}{n}$, $n$ is the number of the neurons or length of the rows of each layer weight matrix.
    \item Linear function: $pr_{j}=\frac{f_{p}}{\sum_{j}f_{p}}$
    \item Half-Gaussian function: $pr_{j}=\frac{\sqrt{2}}{\sigma\sqrt{\pi}}\text{exp}\left({\frac{-f_{p}^{2}}{2\sigma^{2}}}\right)$ 
    \item Softmax function: $pr_{j}=\frac{\text{exp}(f_{p})}{\sum_{j}\text{exp}(f_{p})}$
\end{itemize}
The simplest of these QP functions is uniform or constant function but is not appealing because it is independent of the QE and therefore ignores the random quantization proposition. The most intriguing of all is the half-Gaussian function because of the extra parameter ($\sigma$), which can be learned but is more complicated. The linear and softmax functions have been found to yield nearly the same performance, but the former is simpler to implement. {Accordingly}, in this work, we use the linear function because it balances between performance and simplicity.

\subsection{Low-bit Activation Function}
The inputs to convolutional and fully connected layers are often the outputs of the previous layers' activations. In many low-bit  DNNs designs, the activation layer is often left in its full-precision. However, quantizing the activation layer is crucial in replacing the floating-point operations with more efficient binarization. The conventional activation functions such as \textit{``Relu"} {may not be} suitable for low-bit  DNNs \cite{zhang2018shufflenet}. Therefore, the activations are quantized from 32-bit($\mathbf{u}_{32}$) to $k-\text{bit}$ according to the function
\begin{equation}\label{quant_active}
    \mathbf{W}_{b}=\frac{round\left(\left(\mathbf{W}_{32}-x\right)\cdot\left(2^{k}-1\right)/(y-x)\right)}{\left(2^{k}-1\right)}
\end{equation}
where $\mathbf{W}_{32}$ is the floating-point activation bounded by the input dimension $(x, y)$ and $k=2$. The activations are not stochastically quantized because, unlike in weights, the activations do not have learning parameters. 

\subsection{Model Training and Inference}\label{training}
 \subsubsection{SLP-DNet and Classically Quantized SLP-DNet}
{The SLP-DNet is trained the same way as its corresponding classically quantized versions based on binary and ternary bits (SLP-DBNet and SLP-DTNet).} Each PUM block contains three main components and is trained block-wise for \textit{k}-th number of iterations. Similarly, APM is trained for \textit{r}-th iterations, and the number of training iterations of the PUM and APM may not necessarily be equal. The PUM is trained for 20 iterations and the APM for 10 iterations. We modify the learning rate by a factor $\alpha \in \mathbb{R}^{+}$ for every training step to improve the training efficiency using a stochastic gradient descent algorithm with Adam optimizer \cite{shalev2014understanding}.\par 
\subsubsection{Stochastic Quantized SLP-DNet {(SLP-DSQNet)}}
The SLP-DNet training is slightly different from that of SLP-DNet. The training is summarized in four stages: stochastic weight matrix division, forward propagation, backward propagation, and parameter update. Given QR, the weight matrix is partitioned into a quantization group and a full-precision group using Algorithm \ref{algorithm_SQ_Division}. A hybrid weight is then formed containing the quantized and the real-valued weights, and it provides a better gradient direction than pure quantized weights. If $\Tilde{\mathbcal{W}}_{qf}$ is the composite weight matrix, the weight update with respect to the composite gradients is given by
$\mathbcal{W}^{r+1}=\mathbcal{W}^{r}-\eta\frac{\partial{\mathbcal{L}}}{\partial{\Tilde{\mathbcal{W}}_{qf}^{r}}}$. We train the network with different QRs, which are fixed for all the training iterations and inference.\par
The learning is performed in an unsupervised fashion in which the loss function is the Lagrangian function's statistical mean over the training batch. During the inference, a feed-forward pass is performed over the whole layers using the learned Lagrangian multipliers to compute the precoding vector using (\ref{optima_prec_rel}) and (\ref{robust_optimal}) for nonrobust and robust SLP formulations. Note that except where necessary stated, the training SINR is drawn from a random uniform distribution to enable learning across a wide range of SINR values.

\subsection{Computational Complexity Analysis}
This subsection presents the analytical evaluations of the computational costs of the proposed SLP-DSQNet precoding schemes and compares them with SLP-DNet, the conventional BLP, and the SLP optimization-based methods. The complexities are computed in terms of the number of real arithmetic operations involved. To derive the analytical complexity of the optimization-based SLP, we first convert the second-order cone programming (SOCP) (\ref{P_relaxed2}) into standard linear programming (LP) as follows
\begin{equation}\label{P_relaxed21}
    \begin{aligned}
    & \underset{\mathbf{\{u_{1}\}}}{\text{min}}
    & & {\norm{\mathbf{u}_{1}}_{2}^2} \\
    & \text{s.t.}
    & & |{\boldsymbol{\Phi}_{i}^{T}\boldsymbol{\Upsilon}\mathbf{u}_{1}}|\leq\bar{b}\ ,\ \forall {i}.
    \end{aligned}
\end{equation} 
where $\bar{b}=\left(\boldsymbol{\Phi}_{i}^{T}\boldsymbol{\Upsilon}\mathbf{u}_{1}-\sqrt{\Gamma_{i}n_{0}}\right){\text{tan}{\theta}}$.
To convert (\ref{P_relaxed21}) to its equivalent LP, we introduce new optimization variable
\begin{equation} \label{P_complexity}
    \begin{aligned}
    & \underset{\mathbf{\{x\}}}{\text{min}}
    & & \mathbf{d}^{T}\mathbf{x} \\
    & \text{s.t.}
    & & \mathbf{d}_{k}^{T}\mathbf{x}\leq{-\text{tan}\theta\sqrt{\Gamma_{i}n_{0}}}\ ,\ \forall {i}
    \end{aligned}
\end{equation}
where $\mathbf{d}=[0\ \ \mathbf{u}_{1}^{T}]^{T} \in \mathbb{R}^{(2M+1)\times{1}}$, $\mathbf{x} =[1\ \mathbf{u}_{1}]^{T}\in \mathbb{R}^{(2M+1)\times1}$, $\mathbf{d}_{k}=\left[
\left|\boldsymbol{\Phi }_{i}^{T}\boldsymbol{\Upsilon }\mathbf{u}_{1}\right|\ \ -\boldsymbol{\Phi}_{i}^{T}\tan \theta
\right]^{T}\in \mathbb{R}^{(2M+1)\times1}$ and $\mathbf{U}=[\mathbf{u}_{11}, \cdots,\mathbf{u}_{1K}]; \ \ \forall{i=1,\cdots,K}$.\par
Given the optimal target accuracy, $\epsilon>0$, the complexity of solving convex optimization via IPM is characterized by the formation (${C}_\text{form}$) and factorization (${C}_\text{fact}$) of the matrix coefficients with $\bar{n}$ linear equations having $\bar{n}$ unknowns and is given by \cite{wang2014outage}
\begin{equation}\label{cost1}
    {C}_\text{total}=\left({C}_\text{form}+{C}_\text{fact}\right)\times{\ln\left(\frac{1}{\epsilon}\right)}\sqrt{\sum_{j=1}^{M_\text{lc}}{Q}_{j}+2M_\text{sc}}
\end{equation}
where $Q$ represents the constraint's dimension, $M_\text{lc}$ and $M_\text{sc}$ denote the numbers of linear inequality matrix and second order cone (SOC) constraints, respectively. Therefore, the overall complexity is
\begin{multline}\label{total_cost}
    {C}_\text{total}=\underbrace{\left[\underbrace{\bar{n}\sum_{j=1}^{M_\text{lc}}Q_{j}^{3}+\bar{n}^{2}\sum_{j=1}^{M_\text{lc}}Q_{j}^{2}}_\text{due to $M_\text{lc}$} +\underbrace{\bar{n}\sum_{j=1}^{M_\text{sc}}Q_{j=1}^{2}}_\text{due to $M_\text{sc}$}+\bar{n}^{3}\right]}_{{C}_\text{form}+{C}_\text{fact}}\times\\
    {\ln\left(\frac{1}{\epsilon}\right)}\sqrt{\sum_{j=1}^{M_\text{lc}}{Q}_{j}+2M_\text{sc}}.
\end{multline}
It can be observed that (\ref{P_complexity}) has $K$ constraints with dimension $2M+1$. Therefore, using (\ref{total_cost}), the total computational cost is obtained as
\begin{equation}
    \scalebox{0.96}{${C}_\text{total}=\sqrt{2M+1}\left[\bar{n}(2M+1)+\bar{n}(2M+1)^{2}+\bar{n}^{3}\right]\ln{\left(\frac{1}{\epsilon}\right)}$.}
\end{equation}
By following similar principles and steps above, we can obtain the complexities of the robust SLP and the conventional BLP schemes.\par
On the other hand, to determine the complexities of our proposed precoders, we first evaluate the complexities of the learning modules (PUM and APM) in terms of arithmetic operations involved. For PUM, there are three convolution blocks. The feature map determines the arithmetic operations for a convolution layer and is given by the number of multiplications and additions involved in the convolution operation. The number of operations in a given convolutional layer is
\begin{equation}\label{CNN_COMP}
    C_\text{conv}=\left(c_\text{in}k_\text{f}^{2}+(c_\text{in}k_\text{f}^{2}-1)+1\right)c_{out}{N}_\text{w}{N}_\text{h}
\end{equation}
where $N_\text{h}$, $N_\text{w}$, $k_\text{f}$, $C_\text{in}$ and $C_\text{out}$ denote the height, width of the input layer tensor, filter size, number of input and output channels, respectively.
It is important to note that only the first and second convolutions are quantized, while the last convolution is not to avoid losing essential features of the output precoder. Since in our proposed approach, the layer weight matrix contains both floating points and quantized entries, then the quantization approximation of convolution has $\frac{1}{32}\left(c_\text{in}k_\text{f}^{2}{N}_\text{w}{N}_\text{h}c_{out}\right)\times QR$ binary operations and $\left(c_\text{in}k_\text{f}^{2}{N}_\text{w}{N}_\text{h}c_{out}\right)\times (1-QR)$ non binary operations based on (\ref{CNN_COMP}). 
Using these expressions, we obtain the generic complexity of the PUM as
\begin{multline}\label{comp_PUM}
C_\text{PUM}=\underbrace{\frac{1}{32}\sum^{L}_{l=1}N_\text{h}^{[l-1]}N_\text{w}^{[l-1]}\left[C_\text{in}^{[l-1]}f^{[l]2}\right]C_\text{out}^{[l]}(QR)}_\text{binary operations}+\\
\underbrace{\sum^{L}_{l=1}N_\text{h}^{[l-1]}N_\text{w}^{[l-1]}\left[C_\text{in}^{[l-1]}f^{[l]2}\right]C_\text{out}^{[l]}(1-QR)}_\text{floating point operations}.
\end{multline}
Similarly, the APM's complexity is determined by the cost of the feed-forward pass of the shallow CNN, as shown in Table \ref{tab:proximity_barrier_NN} and the \textit{`log'} barrier that form the barrier term. 
\begin{multline}\label{comp_APM}
    C_\text{APM}=\sum^{L_\text{cv}}_{l=1}N_\text{h}^{[l-1]}N_\text{w}^{[l-1]}\left[C_\text{in}^{[l-1]}f^{[l]2}\right]C_\text{out}^{[l]}+\\
    \sum^{L_\text{fc}}_{j=1}\left(2N_\text{in}^{[j-1]}+1\right)N_\text{out}^{[i]}+\\C_\text{log-barrier}
\end{multline}
where $L_\text{cv}$ and $L_\text{fc}$ are the number of convolution and fully connected layers, respectively. Based on the matrix/vector multiplications, the square absolute and $l_{2}$ norm values, the number of arithmetic operations involved in computing the terms in the \textit{`log'} barrier functions for SLP-DNet and robust SLP-DNet are obtained as $4M^{2}K+2MK+K$ and $8M^{2}K+4MK+6K$, respectively.\par
Finally, we use the information in Tables \ref{tab:proximity_barrier_NN} and \ref{tab:auxiliary_NN} along with (\ref{comp_PUM}) and (\ref{comp_APM}) to obtain the complexity of SLP-DSQBNet as follows
\begin{multline}\label{comp_SLP-DSQBNet}
    C_\text{SQB}=2704K^{2}M+430KM+4M^{2}K-K\\-
    \left[2577K^{2}M+423KM+\frac{7}{8}\right]\times{QR}.
\end{multline}
We can obtain SLP-DSQTNet's complexity from (\ref{comp_SLP-DSQBNet}) by introducing  additional {`0'} state, and this additional bit yields
\begin{multline}\label{comp_SLP-DSQTNet}
    C_\text{SQT}=2704K^{2}M+430KM+4M^{2}K-K-\\
    \left[2433K^{2}M+\frac{783}{2}KM+\frac{7}{8}\right]\times{QR}
\end{multline}
We observe that by substituting $QR=0$ in (\ref{comp_SLP-DSQBNet}) or (\ref{comp_SLP-DSQTNet}), we can obtain the complexity of SLP-DNet. Similarly, the complexities of SLP-DBNet and SLP-DTNet are also found by substituting $QR=1$ in (\ref{comp_SLP-DSQBNet}) and (\ref{comp_SLP-DSQTNet}), respectively. Table \ref{tab:complexity} shows the complexities of the proposed and benchmarks precoding schemes. For illustration, we use the case of symmetry, where $(M=K=\bar{n})$, and show that our proposals have a considerably lower computational complexity of $\mathbcal{O}(n^{3})$. In contrast, the optimization-based SLP and conventional BLP methods have $\mathbcal{O}(n^{6.5})$ and $\mathbcal{O}(n^{7.5})$ computational complexities, respectively. While our proposed schemes have the same order of complexity as SLP-DNet (see Table \ref{tab:complexity-table}), the number of arithmetic operations involved in their computations is lower than that of the SLP-DNet due to the presence of binary operations.

\begin{table*}[!t]
\resizebox{0.92\textwidth}{!}{\begin{minipage}{\textwidth}
\caption{Complexity analysis of proposed SLP-DSQNet and benchmark SLP schemes.}
\label{tab:complexity-table}
\centering
\begin{tabular}{l|l|l}
    \hline
    Method  & Arithmetic Operations (term; $\bar{n}=\mathcal{O}(2KM)$) & Complexity Order ($\bar{n}=M=K$)\\
    \hline
    \hline
     Conventional BLP & $\sqrt{(4M+K+2)}\left[\bar{n}(2M+1)+\bar{n}(2M+1)^{2}+\bar{n}(K+1)^{2}+\bar{n}^{3}\right]\ln{\left(\frac{1}{\epsilon}\right)}$ & $\mathcal{O}(n^{6.5})$\\
    \hline
    SLP Optimization-based  &   $\sqrt{2M+1}\left[\bar{n}(2M+1)+\bar{n}(2M+1)^{2}+\bar{n}^{3}\right]\ln{\left(\frac{1}{\epsilon}\right)}$ & $\mathcal{O}(n^{6.5})$\\
    \hline
    SLP-DNet &  $2704K^{2}M+4M^{2}K+430KM-K$ & $\mathcal{O}(n^{3})$\\
    \hline
    SLP-DBNet & $127K^{2}M+4M^{2}K+7KM-K-\frac{7}{8}$ & $\mathcal{O}(n^{3})$\\
    \hline
    SLP-DTNet & $271K^{2}M+4M^{2}K+\frac{77}{2}KM-K-\frac{7}{8}$ & $\mathcal{O}(n^{3})$\\
    \hline
    SLP-DSQBNet & $2704K^{2}M+430KM+4M^{2}K-K-
    \left[2577K^{2}M+423KM+\frac{7}{8}\right]\times{QR}$& $\mathcal{O}(n^{3})$\\
     \hline
    SLP-DSQTNet & $2704K^{2}M+430KM+4M^{2}K-K-
    \left[2433K^{2}M+\frac{783}{2}KM+\frac{7}{8}\right]\times{QR}$& $\mathcal{O}(n^{3})$\\
     \hline
    Robust Conventional BLP &  $\sqrt{2K(2M+1)}\left[\bar{n}K(2M+1)^{3}+\bar{n}^{2}K(2M+1)^{2}+\bar{n}^{3}\right]\ln{\left(\frac{1}{\epsilon}\right)}$ & $\mathcal{O}(n^{7.5})$\\
     \hline
    Robust SLP Optimization-based    &  $\sqrt{2(2M+1)}\left[2\bar{n}K(2M+1)^{2}+\bar{n}^{3}\right]\ln{\left(\frac{1}{\epsilon}\right)}$ & $\mathcal{O}(n^{6.5})$\\
    \hline
    Robust SLP-DNet & 
    $2704K^{2}M+8M^{2}K+432KM+8M^{2}K+6K-2$ & $\mathcal{O}(n^{3})$\\
    \hline
    Robust SLP-DBNet & $127K^{2}MK+8M^{2}K+9KM+6K-\frac{9}{8}$ & $\mathcal{O}(n^{3})$\\
    \hline
    Robust SLP-DTNet & $271K^{2}M+8M^{2}K+\frac{81}{2}KM+6K-\frac{9}{8}$ & $\mathcal{O}(n^{3})$\\
    \hline
    Robust SLP-DSQBNet & $2704K^{2}M+8M^{2}K+432KM+6K-2-
    \left[2577K^{2}M+423KM+\frac{7}{8}\right]\times{QR}$& $\mathcal{O}(n^{3})$\\
     \hline
    Robust SLP-DSQTNet & $2704K^{2}M+8M^{2}K+432KM+6K-2-
    \left[2433K^{2}M+\frac{783}{2}KM+\frac{7}{8}\right]\times{QR}$& $\mathcal{O}(n^{3})$\\
     \hline
    \hline
\end{tabular}
\end{minipage}}
\end{table*}

\begin{table}[!htb]
\renewcommand{\arraystretch}{1.3}
\caption{Simulation settings}
\label{tab:experimental_parameter}
\centering
\begin{tabular}{ll}
    \hline
    Parameters  &  Values\\
    \hline
    \hline
    Training Samples    &  50000 \\
    \hline
    Batch Size (B)   &  200 \\
    \hline
    Test Samples    &  2000 \\
    \hline
    Training SINR range    &  0.0dB - 45.0dB \\
    \hline
    Test SINR range (\textit{i}-th user SINR)    &  0.0dB - 35.0dB\\ 
    \hline
    Optimizer    &  SGD with Adam \\
    \hline
    Initial Learning Rate, $\eta$   &  0.001 \\
    \hline
    Learning Rate decay factor, $\alpha$  &  0.65 \\
    \hline
    Lower bit Activation & bits-width, $k=2$ \\
    \hline
    Number of blocks in the PUM & $B_{l}=2$ \\
    \hline
    Training Iterations in the PUM  per block  &  20 \\
    \hline
    Training iterations for the APM &  10 \\
    \hline
\end{tabular}
\end{table}
\begin{table}[!htb]
\renewcommand{\arraystretch}{1.3}
\caption{Proximity Barrier Term NN Layout}
\label{tab:proximity_barrier_NN}
\centering
\begin{tabular}{l|l}
    \hline
    Layer  &  Parameter, $\text{kernel size}=3\times3$\\
    \hline
    \hline
    Input Layer    &  Input size $(\text{B},\ 1,\ 2M,\ K)$ \\
    \hline
    Layer 1: Convolutional    & Size $(\text{B},20,2M, K^{2})$; zero padding\\
    \hline
    Layer 2: Average Pooling    & Size $((1,\ 1),\ \text{stride}=(1,\ 1))$\\ 
    \hline
    Layer 3: Activation    & Soft-Plus \\ 
    \hline
    Layer 4: Flat
    &  Size $(\text{B}\times40\times K^{2})$ \\
    \hline
    Layer : Fully-connected  & Size$(\text{B}\times40\times K^{2},\ 1)$\\
    \hline
    Layer 5: Activation    & Soft-Plus function\\ 
    \hline
\end{tabular}
\end{table}

\begin{table}[!htb]
\renewcommand{\arraystretch}{1.3}
\caption{An APM NN Architecture}
\label{tab:auxiliary_NN}
\centering
\begin{tabular}{l|l}
    \hline
    Layer  &  Parameter, $\text{kernel size}=3\times3$\\
    \hline
    \hline
    Input Layer    &  Input size $(\text{B},\ 1,\ 2M,\ K)$ \\
    \hline
    Layer 1: Convolutional    & Size $(\text{B},\ 16,\ 2M,\ K)$,\\ & $\text{dilation} = 1$ and unit padding\\
    \hline
    Layer 2: Batch Normalization
    &  $\text{eps}=10^{-6}$, $\text{momentum}=0.1$\\
    \hline
    Layer 3: Activation   &  PReLu/k-bit function\\
    \hline
    Layer 4: Convolutional    & Size $(\text{B},\ 8,\ K,\ 2KM)$,\\ &
    $\text{dilation} = 1$ and unit padding\\
    \hline
    Layer 5: Batch Normalization
    &  $\text{eps}=10^{-6}$, $\text{momentum}=0.1$\\
    \hline
    Layer 6: Activation   & PReLu/k-bit function \\
    \hline
    Layer 7: Convolutional    & Size $(\text{B},\ 1,\ 2KM,\ 1)$,\\ & $\text{dilation} = 1$ and unit padding\\
    \hline
\end{tabular}
\end{table}

\section{Simulation Results and Discussion}\label{results}
\subsection{Simulation Set-up}
\begin{figure}[!t]
    \centering
    \includegraphics[width=3.6in,height=1.90in]{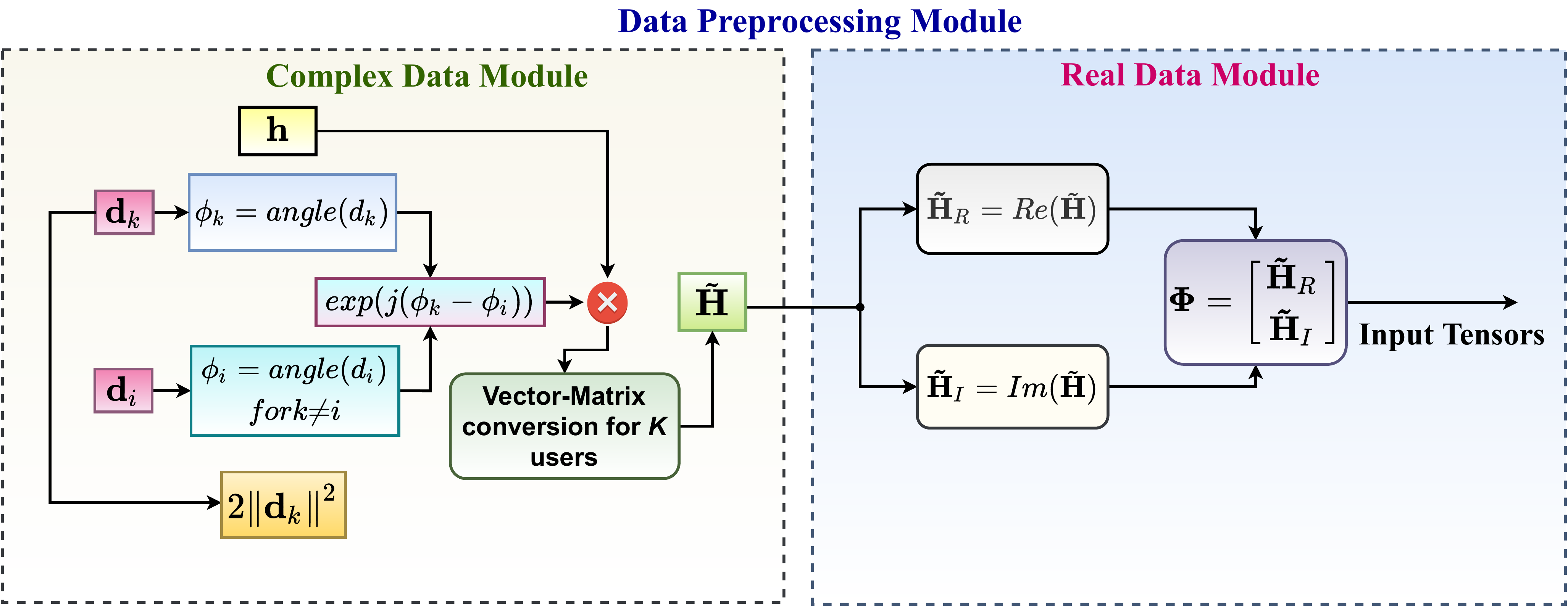}
    \caption{Dataset generating and preprocessing Block.}
    \label{fig:DATASET_Block}
\end{figure}

We consider a downlink situation in which the BS is equipped with four antennas ($M=4$) that serve $K$ single users; and assume a single cell. We obtain the dataset from the channel realizations randomly generated from a normal distribution with zero mean and unit variance. The dataset is reshaped and converted to real number domain using the following expression $\boldsymbol{\Phi} =\begin{bmatrix}
\Tilde{\mathbf{h}}_{Ri}; \ \Tilde{\mathbf{h}}_{Ii}\
\end{bmatrix}$ as summarized in Fig. \ref{fig:DATASET_Block}. The input dataset is normalized by the transmit data symbol so that data entries are within the nominal range, potentially aiding the training. We generate 50,000 training samples and 2000 test samples, respectively. The transmit data symbols are modulated using a QPSK modulation scheme. The training SINR is obtained random from uniform distribution $\Gamma_{\text{train}} \sim \mathcal{U}(\Gamma_\text{low},\,\Gamma_\text{high})$. Stochastic gradient descent is used with the Lagrangian function as a loss metric. A parametric rectified linear unit (\textbf{PReLu}) activation function is used for both convolutional and fully connected layers in a full-precision SLP-DNet and the low-bit activation function (\ref{quant_active}) for SLP-SQDNet. After every iteration, the learning rate is reduced by a factor $\alpha=0.65$ to help the learning algorithm converge faster. The models are implemented in Pytorch 1.7.1 and Python 3.7.8 on a computer with the following specifications: Intel(R) Core (TM) i7-6700 CPU Core, 32.0GB of RAM. Tables 1 summarizes the simulation parameters, while Tables 2 and 3 depict the NN component settings of the SLP-DNet \cite{mohammad2021unsupervised}.

\subsection{Performance Evaluation of QSLP-DNet and SLP-DNet}\label{nonrobust_performance}
{In the following set of results} we compare our proposed quantized DL-based SLP scheme's performance against its corresponding full-precision (SLP-DNet) counterpart's \cite{mohammad2021unsupervised} and other benchmark schemes, such as conventional BLP \cite{zheng2008robust}\cite{bjornson2014optimal} and the optimization-based {SLP} \cite{masouros2015exploiting}. Primarily, we design full low-bit binary and ternary SLP-DNet models (SLP-DBNet and SLP-DTNet), where the real-valued weights and activation are constrained to 1-bit. Similarly, the expressive learning abilities of SLP-DBNet and SLP-DTNet are further enhanced by designing their corresponding low-bit hybrid stochastically quantized versions (SLP-DSQBNet and SLP-DSQTNet), where part of the weight matrix is quantized to a lower bit, while the remaining is left in its 32-bit floating-point precision. The resulting weight matrix is a hybrid containing both binary and real-valued entries with the activations all reduced to 2-bit according to (\ref{quant_active}).\par

\begin{figure}[!t]
    \centering
    \includegraphics[width=2.7in,height=2.4in]{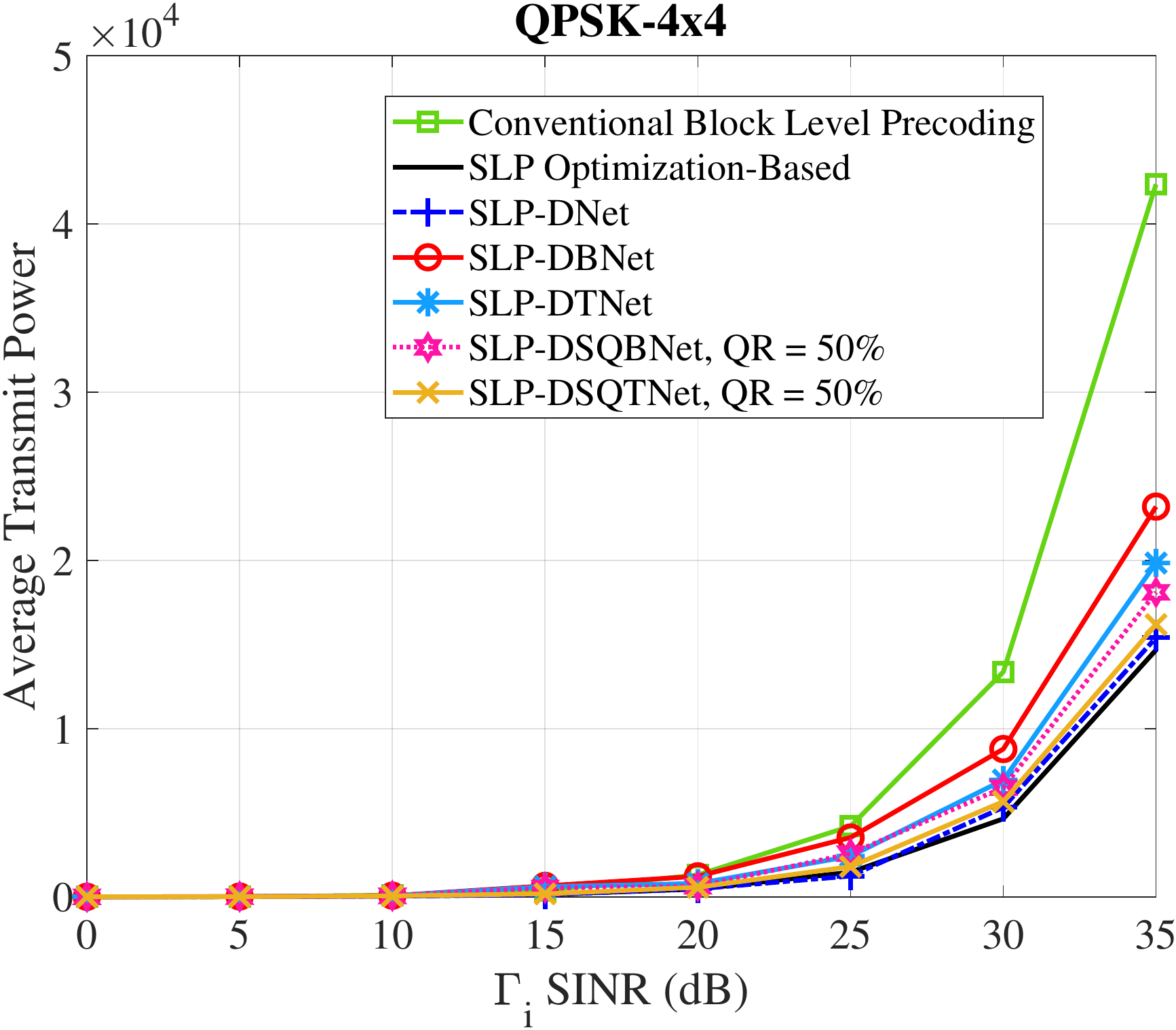}
    \caption{Transmit Power vs SINR averaged over 2000 test samples for Conventional Block Level Precoding, SLP optimization-based and nonrobust quantized learning-based SLP solutions, $M=4$, $K=4$ and $QR=50\%$.}
    \label{fig:POWER_vs_SINR_SQ_Nonrobust}
\end{figure}

The performances of SLP-DBNet, SLP-DTNet, SLP-DSQBNet, SLP-DSQTNet {for $QR=0.5$} against SLP-DNet and other benchmark precoding schemes (conventional BLP, SLP optimization-based) are shown in Fig. \ref{fig:POWER_vs_SINR_SQ_Nonrobust}. It can be observed that both SLP-DBNet and SLP-DTNet have higher transmit power than the SLP optimization-based and SLP-DNet schemes. Therefore, SLP optimization-based and SLP-DNet solutions require less power to transmit the same amount of data symbols than SLP-DBNet and SLP-DTNet. The {loss} in performance is expected because some information is lost during feed-forward weight/input convolutions due to quantization and the inhomogeneous nature of the quantization errors.\par
Furthermore, a closer examination of Fig. \ref{fig:POWER_vs_SINR_SQ_Nonrobust} reveals that the SLP-DSQBNet and SLP-DSQTNet offer less transmit power than their corresponding full binary and ternary versions. Our simulation also shows that learning by stochastic quantization results in the performance close to the full-precision learning model (SLP-DNet) with a significant model size reduction (memory savings at the inference), as we shall see later. We argue that the decrease in the available transmit power at the BS in this scenario is because not all the weights matrix rows are quantized at once. The quantization error is used to direct the gradient descent towards the best local minima during training. Accordingly, we find that at 30dB, the performance of SLP-DBNet and SLP-DTNet falls by 58\% and 35\% of the SLP optimization-based solution, respectively. On the other hand, the performance gaps of SLP-DSQBNet, SLP-DSQTNet, and SLP-DNet are 22.2\%, 9.62\%, and 5\% of the SLP optimization-based solution, respectively. Therefore, while the fully quantized model's accuracy is significantly low, the stochastically hybrid quantized counterparts and full-precision models' accuracy is within $88\%-96\%$ of the optimal solution.\par

\begin{figure*}
\hspace{5mm}%
\subfloat[Transmit Power vs SINR averaged over 2000 test samples for conventional, SLP optimization-based and robust quantized learning-based SLP solutions under $M=4$, $K=4$ and $\varsigma^{2}=0.0002$\label{fig:POWER_vs_SINR_SQ_Robust}]{%
  \includegraphics[width=2.8in,height=2.4in]{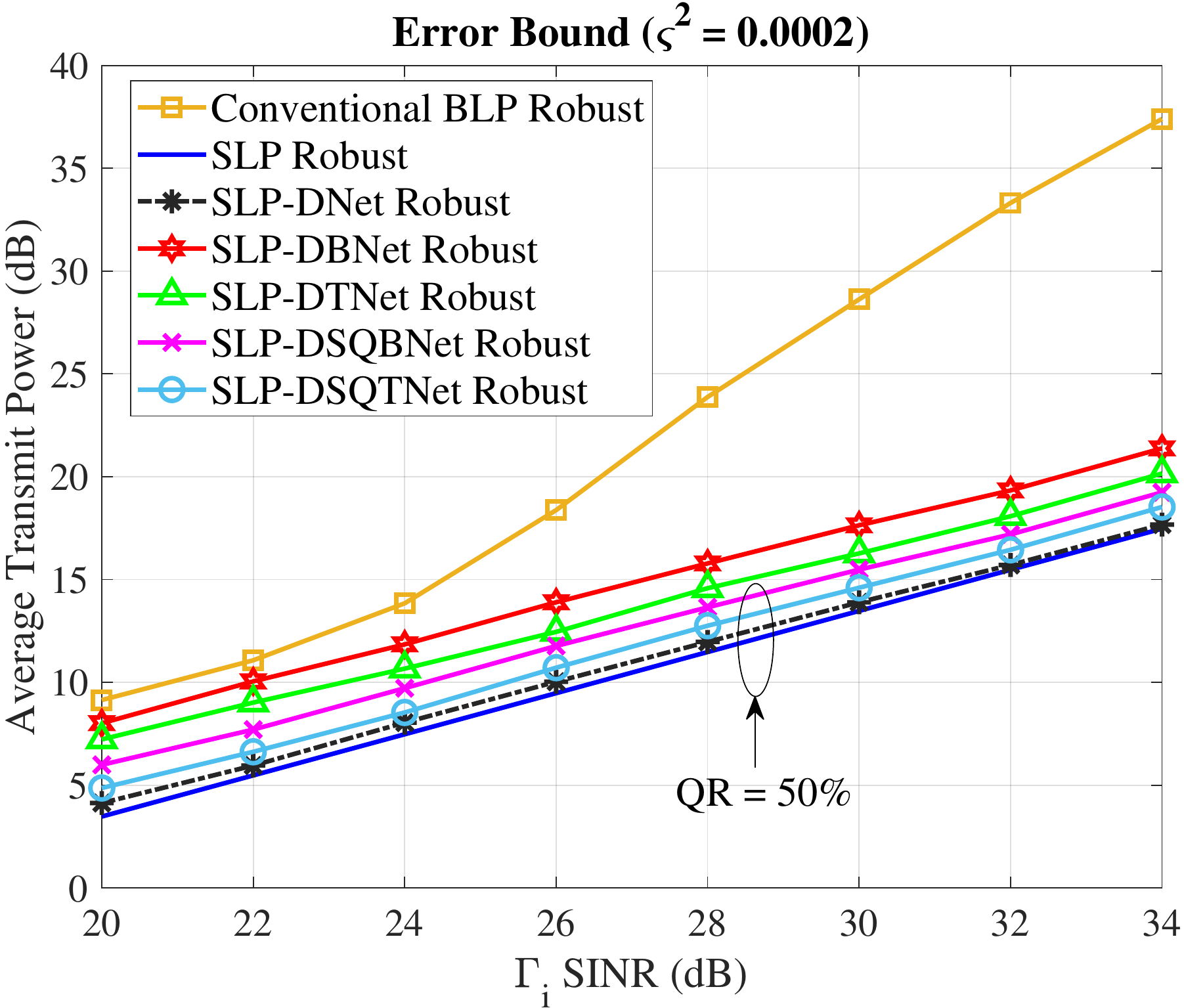}%
}
\hspace{28mm}%
\subfloat[Transmit Power vs Error-bound for Conventional BLP, robust SLP optimization-based and robust qunatized learning-based SLP solutions under $M=4$, $K=4$ and $QR=50\%$\label{fig:POWER_vs_ERRORBOUND}]{%
  \includegraphics[width=2.8in,height=2.4in]{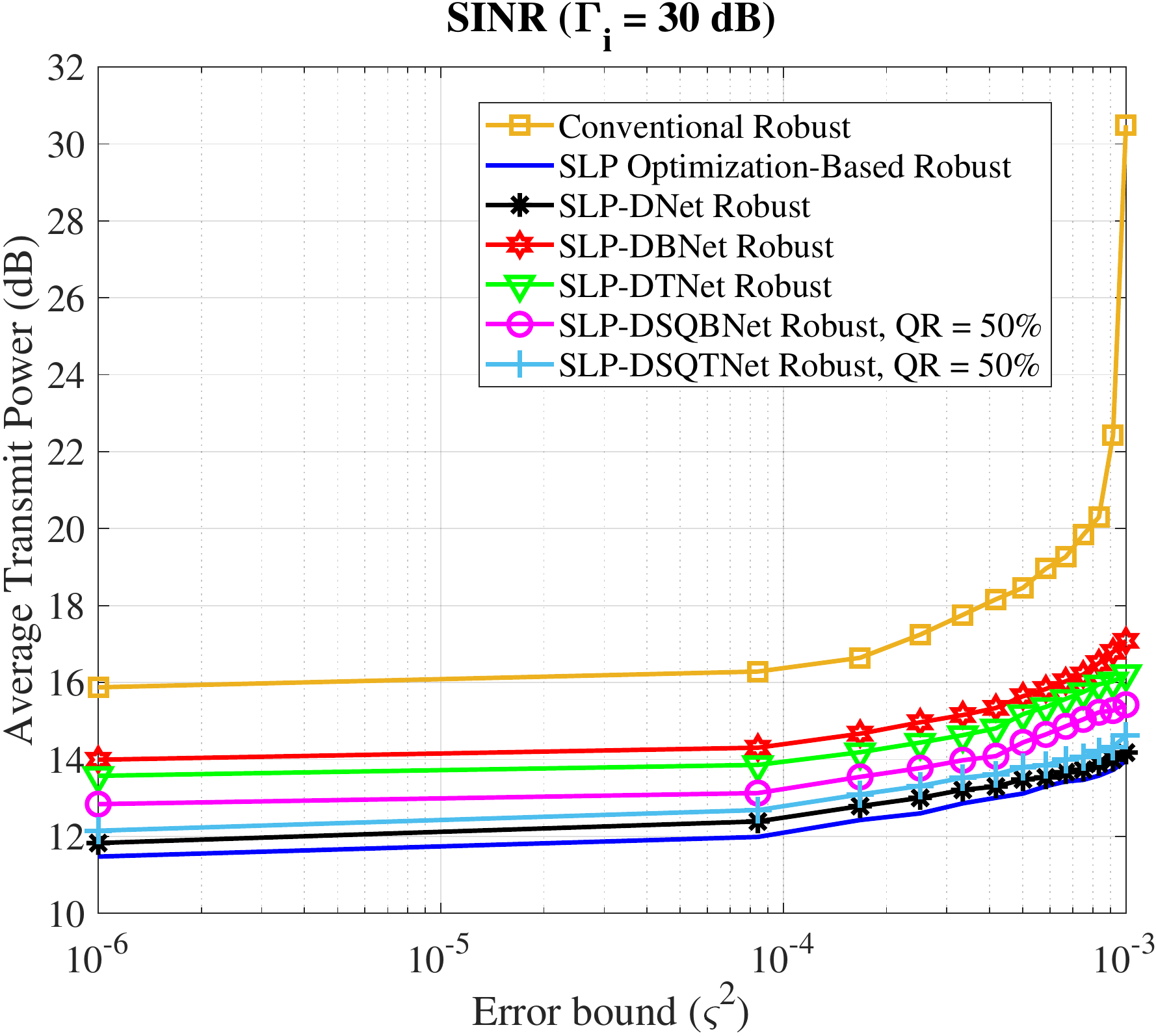}%
}
\caption{Performance evaluation of Robust formulation for Conventional BLP, SLP Optimization-based and SLP Learning-based schemes.}
\label{fig:POWER_vs_SINR_CHANNEL_ERROR}
\end{figure*}

\subsection{Performance Evaluation of Robust SLP-SQDNet}\label{robust_performance}
Figs. \ref{fig:POWER_vs_SINR_SQ_Robust} and \ref{fig:POWER_vs_ERRORBOUND} compare the performances of SLP-SQDNet and the traditional CSI-robust precoder for the $4\times4$ MISO system evaluated at $\varsigma^{2}={10}^{-4}$. Fig. \ref{fig:POWER_vs_SINR_SQ_Robust} depicts how the average transmit power increases with the $SNR$ thresholds, for CSI error bounds $\varsigma^{2}={10}^{-4}$ and $QR=50\%$. The robust SLP optimization-based is observed to show a significant power savings of more than 60\% compared to the robust conventional BLP. Similarly, the proposed unsupervised learning-based precoders portray similar transmit power reduction trend. They show considerable power savings of $40\%-58\%$ against the conventional optimization result. While the fully quantized models have demonstrated substantial performance loss compared to SLP-based optimal precoder, SLP-DSQBNet and SLP-DSQTNet offer $90\%-98\%$ striking performance correlation with the SLP optimization-based optimal solutions, respectively.\par 

\begin{figure}[!t]
    \centering
    \includegraphics[width=2.8in,height=2.4in]{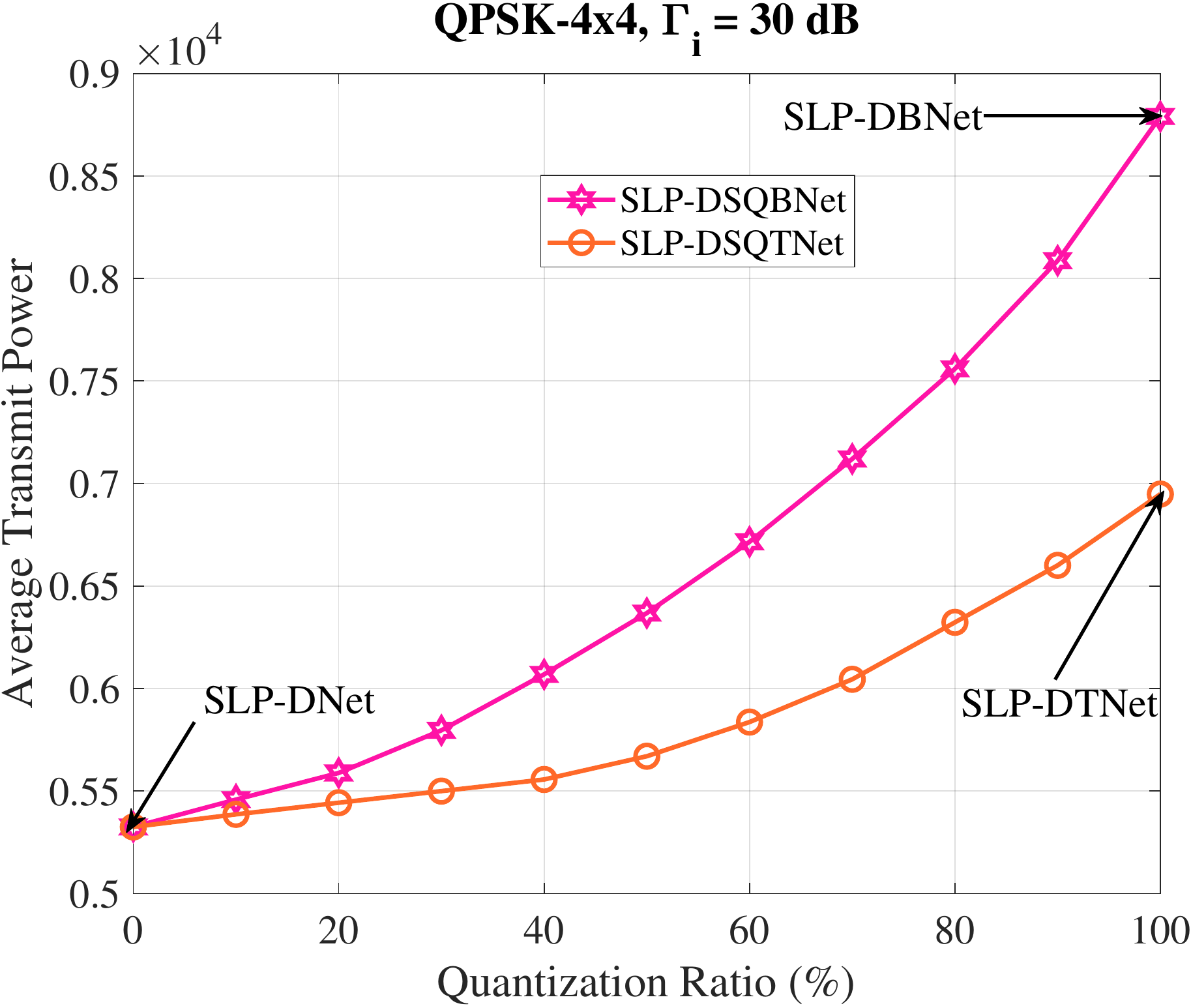}
    \caption{Transmit Power vs Quantization ratio averaged over 2000 test samples for robust SQ SLP-DNet models and full-precision SLP-DNet model under $M=4$, $K=4$ and $\Gamma=30$ dB.}
    \label{fig:SQ_vs_QR_Robust}
\end{figure}

Furthermore, we investigate the effect of the CSI error bounds on the transmit power at 30dB. Fig. \ref{fig:POWER_vs_ERRORBOUND} depicts the variation of the transmit power with increasing CSI error bounds. Moreover, a significant increase in transmit power can be observed where the channel uncertainty lies within the region of CSI error bounds of $\varsigma^{2}=10^{-3}$. Interestingly, like the SLP optimization-based algorithm, by exploiting the CI, the proposed unsupervised learning methods also show a descent or moderate increase in transmit power. To further understand the impact of the $QR$ on the transmit power, Fig. \ref{fig:SQ_vs_QR_Robust} compares the performance of the proposed stochastic quantization learning-based CSI-robust precoders evaluated at 30dB. Like the results obtained for the nonrobust scenario, we also observe a similar trend, where the average transmit power available at the BS required to transmit data symbols increases as more weights and activations are quantized.
\subsection{{Complexity and Memory Evaluation}}
The proposed learning schemes' complexities are examined in two folds: firstly, we compare the number of FLOPs operations involved in our proposed learning methods and those of the benchmark precoding schemes'. Secondly, we evaluate and assess the inference memory requirements of our proposed learning-based precoding techniques.

\subsubsection{Number of FLOPs Operations}
The computational costs of the SLP-DNet {are} obtained from the PUM and the feed-forward convolutions of the CNN that makes up an APM. For the PUM, the dominant {computational cost comes} from computing the proximal barrier term \cite{mohammad2021unsupervised}. It can be seen that both SLP optimization-based algorithm and the proposed learning schemes are feasible for all sets of $M$ BS antennas and $K$ mobile users. However, for conventional BLP, the solution is only feasible for $M \geq K$.\par 

\begin{figure}[!t]
    \centering
    \includegraphics[width=3.2in,height=2.4in]{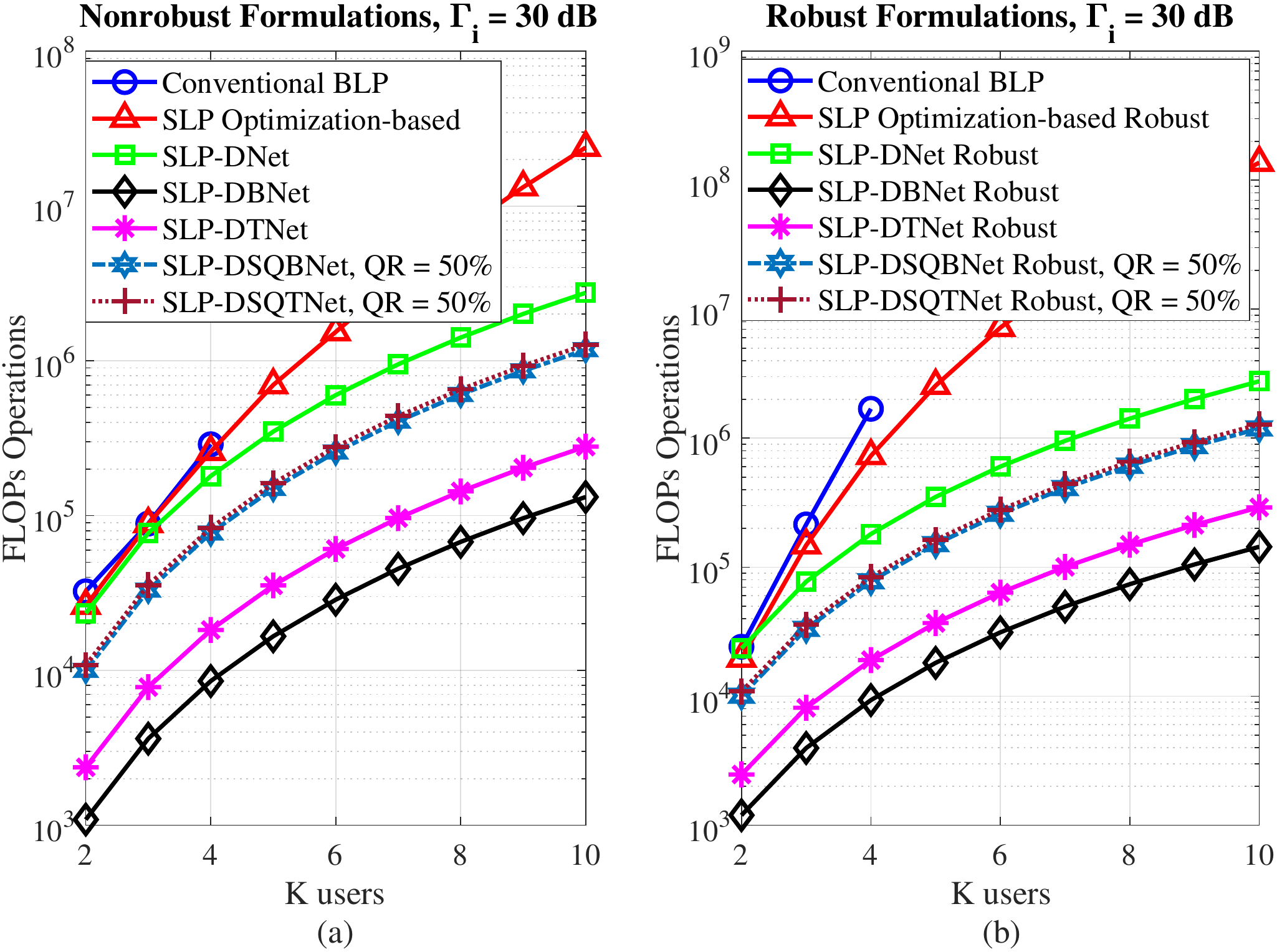}
    \caption{Comparison of FLOPs operations performed for Nonrobust and Robust precoding schemes, i.e, conventional BLP, SLP optimization-based and SLP learning-based models using four BS antennas ($M=4$) and $QR=50\%$.}
    \label{fig:Execusion_time}
\end{figure}

\begin{figure*}
\hspace{5mm}%
\subfloat[Transmit Power vs Quantization Ratio averaged over 2000 test samples for SLP optimization-based and SQ SLP-DNet models for nonrobust formulations, $M=4$, $K=4$ and $\Gamma_{i}=30$ dB\label{fig:SQ_vs_QR_Nonrobust}]{%
  \includegraphics[width=2.8in,height=2.4in]{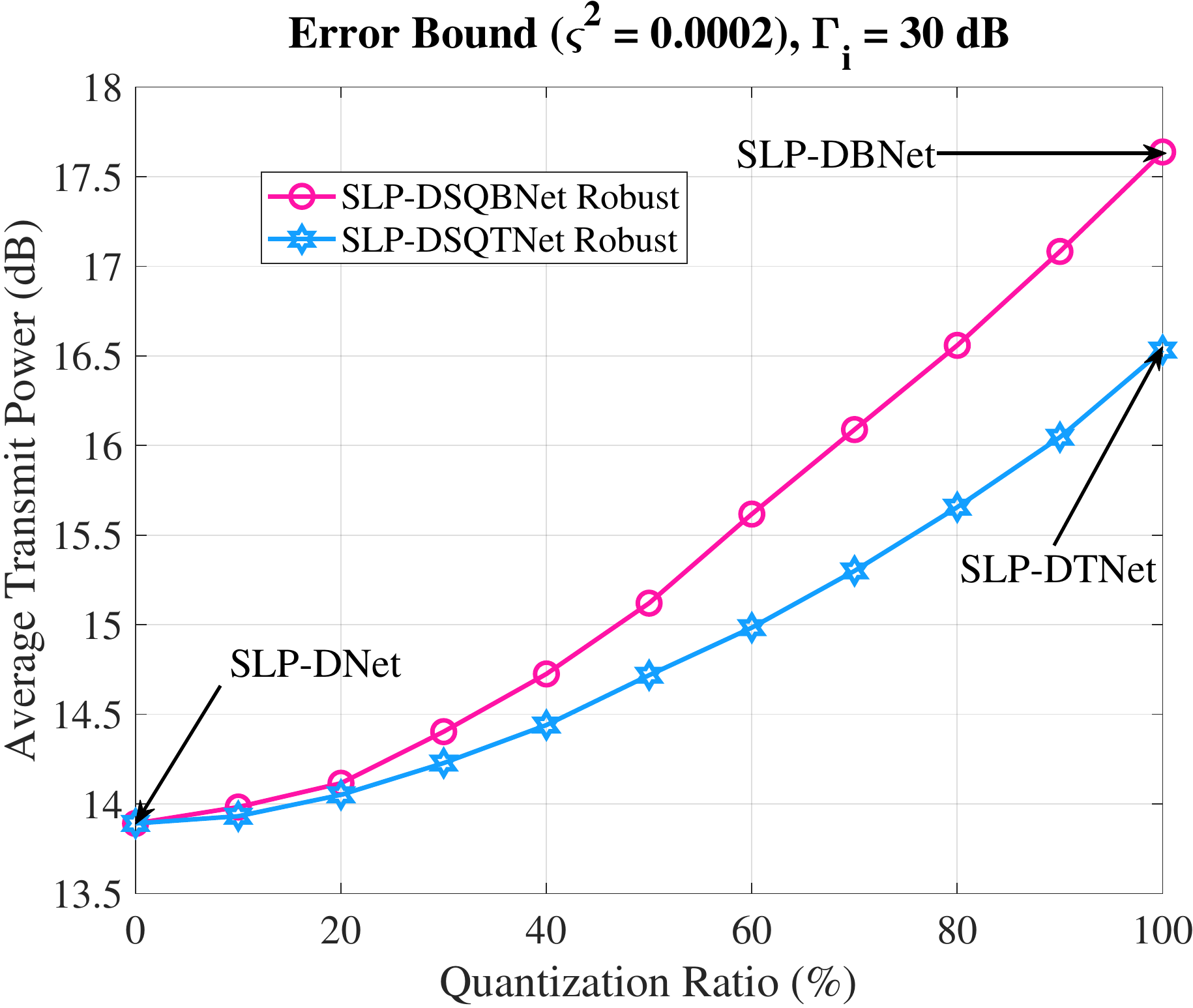}%
}
\hspace{25mm}%
\subfloat[Memory requirement at Inference vs Quantization ratio for SLP-based Full-precision SLP-DNet, SLP-DSQBNet and SLP-DSQTNet under $M=4$, $K=4$ and $\Gamma_{i}=30 \text{dB}$\label{fig:Memory_vs-QR}]{%
  \includegraphics[width=2.8in,height=2.4in]{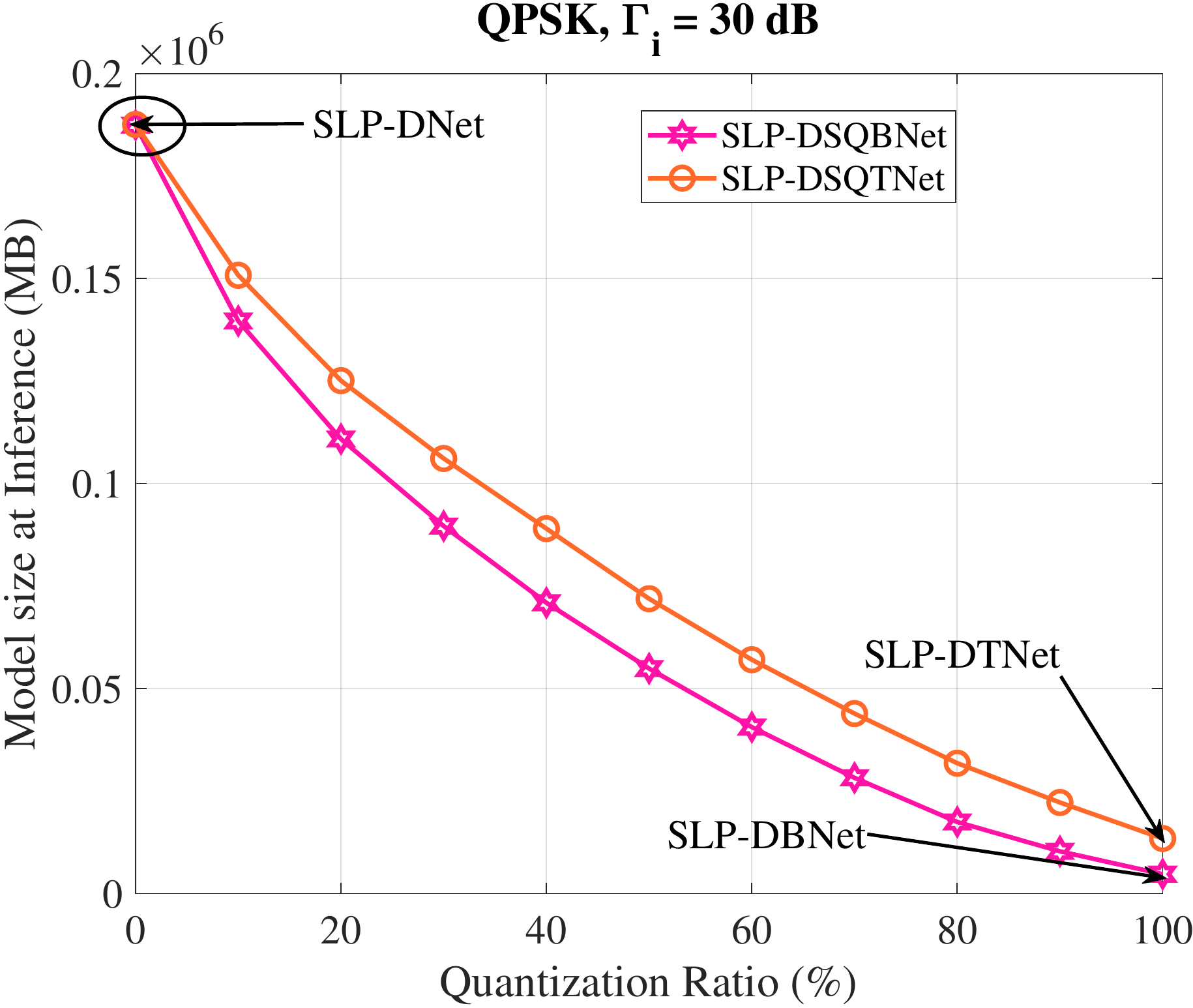}%
}
\caption{Average power and inference memory requirement vs quantization error of the proposed learning-based precoding schemes.}
\label{fig:POWER_MEMORY_VS_QR}
\end{figure*}

\begin{figure}[!t]
    \centering
    \includegraphics[width=2.8in,height=2.4in]{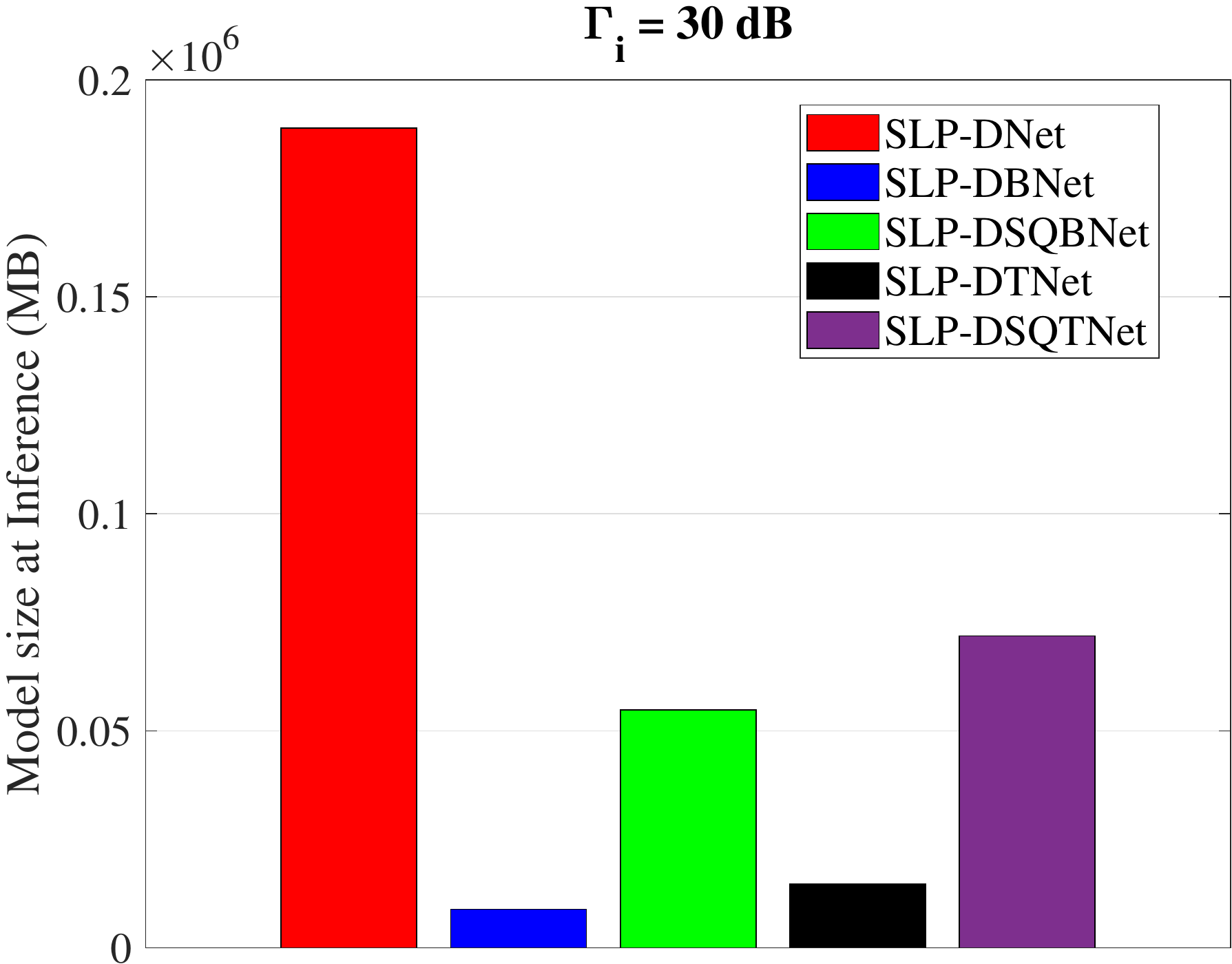}
    \caption{Memory requirement at Inference for Full-precision SLP-DNet, SLP-DBNet, SLP-DTNet, SLP-DSQBNet and SLP-DSQTNet under $M=4$, $K=4$, $\Gamma_{i}=30 \text{dB}$ and $QR=50\%$.}
    \label{fig:Memory_barchart}
\end{figure}

Fig. \ref{fig:Execusion_time}(a) shows the number of FLOPs operations of the proposed unsupervised learning solutions per symbol for nonrobust formulations. The dominant operations involved in SLP-DNet at the inference are matrix-matrix or vector-matrix convolution. The gap in the computational cost between SLP-DNet and SLP optimization-based methods increases with the growing number of mobile users. For example, we find that the complexity of SLP-DNet is $\sim10\times$ lower than SLP optimization-based at $K=10$, while that of SLP-DSQBNet and SLP-DSQTNet are $\sim20\times$ much lower due to the presence of binary operations. Furthermore, SLP-DBNet and SLP-DTNet offer an additional computational complexity reduction than SLP-DSQBNet and SLP-DSQTNet because binary bit-wise operations replace the entire MACs calculations in the for-ward pass. It is important to recall that SLP-DTNet outperforms SLP-DBNet in all scenarios. However, we observe that SLP-DTNet is slightly slower than SLP-DBNet, and this is due to the additional {`0'} binary state introduced in the former. We also note that the {advantages} of the SLP-DBNet and SLP-DTNet are further enhanced via stochastic quantization but at the expense of small additional complexity overhead. The same trend is also observed in the case of a robust channel scenario, as shown in Fig. \ref{fig:Execusion_time}(b).\par 
Accordingly, we can deduce that while fully binarized DNN could offer significant training and inference accelerations, it could otherwise lead to significant performance degradation. However, quantizing the weight matrix via a stochastic channel selection based on the quantization error leads to the improved received power. Therefore, we can conclude that the results in Figs. \ref{fig:Execusion_time}(a) and \ref{fig:Execusion_time}(b) demonstrate that the proposed quantized DL-based SLP solutions offer a good trade-off between the performance and computational complexity.

\begin{table*}[!t]
\renewcommand{\arraystretch}{1.2}
\caption{Inference memory utilization}
\label{tab:complexity}
\centering
\begin{tabular}{lllllll}
    \hline
    \hline
    Models & Weights & Activations  & Memory  & Memory &\\
    & & & usage (MB) & saving & \\
    \hline
    \hline
    SLP-DNet & $(32-\text{bit})\in\mathbb{R} $ & $(32-\text{bit})\in\mathbb{R}$ & 0.1898 & $-$\\
    \hline
    SLP-DBNet & $\{-1,+1\}$ & $\{-1,+1\}$ & 0.0089 & $21.33\times$\\
    \hline
    SLP-DTNet & $\{-1, 0, +1\}$ & $\{-1,+1\}$ & 0.0146 & $13\times$\\ 
    \hline
    SLP-DSQBNet & $\{-\beta_\text{qf}, \beta_\text{qf}\}$ & $\{-\beta_{2-bit}, \beta_{2-bit}\}$ & 0.0548 & $3.46\times$\\ 
    \hline
    SLP-DSQTNet  & $\{-\beta_\text{qf}, 0, \beta_\text{qf}\}$ & $\{-\beta_{2-bit}, \beta_{2-bit}\}$ & 0.0719 & $2.64\times$\\
    \hline
    \hline
\end{tabular}
\end{table*}

\subsubsection{Model Size and Memory Utilization}
Generally, GPU can speedup the offline training of DNNs. However, most modern GPUs are memory-constrained (e.g.GTX 980: 4GB, Tesla K40: 12GB, Tesla K20: 5GB and GTX Titan X: 12GB)\cite{cuda}. Practically, the size of the DNN is often bounded by the available memory. Therefore, it is beneficial to estimate the memory requirements of the DNN at the inference. Likewise, the actual memory utilization also depends on the implementation. Here, we examine and analyze the memory utilization of full-precision SLP-DNet and its corresponding quantized versions at inference. By memory utilization, we refer to the model size at the testing phase. For this analysis, we adopt the approach presented in \cite{bethge2019binarydensenet} to calculate the inference memory utilization as the summation of 32-bit times the number of floating-point parameters and 1-bit times the number of binary parameters. Mathematically, this can be expressed as $\frac{1}{32}W_{b}+ W_{f}$, where $W_{b}$ and $W_{f}$ are the binary and floating-point weights, respectively.\par
Fig. \ref{fig:SQ_vs_QR_Nonrobust} shows the average transmit power vs quantization ratio {(i.e. the proportion of weights that are quantized)} at 30dB SINR. The average power at $QR=0$ corresponds to SLP-DNet while $QR=1$ represents the corresponding fully quantized counterparts (SLP-DBNet and SLP-DTNet). Moreover, the transmit power gradually increases as more weights are quantized. It is important to note that for a unit quantization ratio ($QR=1.0$), all the weights are 100\% quantization, where the model could be either a typical binary or ternary. On this note, it is clear that the SLP-DSQTNet offers less transmit power than SLP-SQDBNet. We find that quantizing half of the weights ($QR=50\%$) could guarantee a good performance within $80\%-98\%$ of the full-precision model for both SLP-SQDBNet and SLP-DSQTNet, respectively. To investigate the amount of the memory required at inference with the increase in the quantization ratio, we plot the model size vs QR as depicted in Fig. \ref{fig:Memory_vs-QR}. We find that less memory is required as the quantization moves towards extreme binarization to the right of the QR-axis. It can be seen that the continuous line represents a full-precision SLP-DNet (i.e., $QR=0$), while $QR=1$ represents a fully quantized model.\par
Furthermore, Fig. \ref{fig:Memory_barchart} shows that SLP-DBNet and SLP-DBNet provide considerable memory savings up to $\sim21\times$ and $\sim13\times$ compared to the full-precision SLP-DNet because the extreme quantization reduces the available learning parameters significantly. This brings about a trade-off between performance and model size, which is compensated by hybrid quantization as in SLP-DSQBNet and SLP-DSQTNet. Table \ref{tab:complexity} presents the summary of the inference memory requirements, MACs, and binary operations of different proposed learning implementations. For SLP-DSQBNet and SLP-DSQTNet, the weights are constrained to the following quantization $\{-\beta_\text{qf}, \beta_\text{qf}\}$ and $\{-\beta_\text{qf}, 0, \beta_\text{qf}\}$ while the activations are clipped to $\{-\beta_{2-bit}, \beta_{2-bit}\}$ ${2-\text{bit}}$ quantized values, respectively.  This shows that the hybrid quantization enhances the representational capabilities of the convolutional block.
 
\section{Conclusion}\label{conclusion}
This paper {proposed} a hybrid quantization DNN-based SLP scheme termed (SLP-QSDNet) based on binary and ternary operations for power minimization for a multi-user downlink MISO system. {We proposed} various weight quantization techniques to obtain its corresponding full and partially quantized counterparts. {We showed that} the proposed approach resulted in fast online learning and a significant model size reduction, which could help render the trained model memory-efficient during deployment on the device's edge. {Overall, our proposed approaches provide a scalable tradeoff between performance and complexity in learning-based SLP transmission.}

\bibliographystyle{IEEEtran}
\bibliography{ref}

\end{document}